\documentclass[usenatbib]{mn2e}
\usepackage{natbib}
\usepackage{graphicx}
\usepackage{epsfig}
\usepackage{verbatim}
\title[Lyman Break Galaxies at $z=4-6$]
{Lyman Break Galaxies at $z=4-6$ in cosmological SPH simulations}

\author[C. Night, K. Nagamine, V. Springel, and L. Hernquist]
{C.~Night$^1$\thanks{Email: cnight@cfa.harvard.edu}
 K.~Nagamine$^2$\thanks{Email: knagamine@ucsd.edu}
 V.~Springel,~$^3$\thanks{Email: volker@mpa-garching.mpg.de} and
 L.~Hernquist$^1$\thanks{Email: lars@cfa.harvard.edu}
   \vspace{0.3cm}\\ 
$^{1}$Harvard-Smithsonian Center for Astrophysics, 
60 Garden Street, Cambridge, MA 02138, U.S.A. \\
$^{2}$University of California, San Diego, 
Center for Astrophysics \& Space Sciences, 9500 Gilman Dr., 
La Jolla, CA 92093-0424, U.S.A. \\
$^{3}$Max-Planck-Institut f\"{u}r Astrophysik, 
        Karl-Schwarzschild-Stra\ss{}e 1, 85740 Garching bei 
        M\"{u}nchen, Germany}

\newcommand{\Ly}{\lambda_{Ly}}
\newcommand{\hc}{h_{70}}
\newcommand{\kmsmpc}{{\rm\ km\ s^{-1}\ Mpc^{-1}}}
\newcommand{\Mstar}{M_{\star}}
\newcommand{\hinv}{{\hc^{-1}}}
\newcommand{\Msun}{M_{\odot}}
\newcommand{\Lsun}{L_{\odot}}
\newcommand{\himsun}{\hinv{\Msun}}
\newcommand{\himpc}{\hinv {\rm Mpc}}

\begin{document}

\maketitle

\label{firstpage}


\begin{abstract}

  We perform a spectrophotometric analysis of galaxies at redshifts
  $z=4-6$ in cosmological SPH simulations of a $\Lambda$ cold dark
  matter ($\Lambda$CDM) universe. Our models include radiative cooling
  and heating by a uniform UV background, star formation, supernova
  feedback, and a phenomenological model for galactic winds.
  Analysing a series of simulations of varying boxsize and particle
  number allows us to isolate the impact of numerical resolution on
  our results.  Specifically, we determine the luminosity functions in
  $B$, $V$, $R$, $i'$, and $z'$ filters, and compare the results with
  observational surveys of Lyman break galaxies (LBGs) performed with 
  the {\it Subaru} telescope and the {\it Hubble Space Telescope}. 
  We find that the simulated galaxies
  have UV colours consistent with observations and fall in the expected
  region of the colour-colour diagrams used by the Subaru group.
  The stellar masses of the most massive galaxies in our largest 
  simulation increase their stellar mass from $\Mstar \sim 10^{11}\Msun$ 
  at $z=6$ to $\Mstar \sim 10^{11.7}\Msun$ at $z=3$. 
  Assuming a uniform extinction of $E(B-V)=0.15$, we also find
  reasonable agreement between simulations and observations in the
  space density of UV bright galaxies at $z=3-6$, down to the
  magnitude limit of each survey.  For the same moderate extinction
  level of $E(B-V)\sim 0.15$, the simulated luminosity functions match
  observational data, but have a steep faint-end slope with $\alpha
  \sim -2.0$.  We discuss the implications of the steep faint-end
  slope found in the simulations. Our results confirm the generic
  conclusion from earlier numerical studies that UV bright LBGs at 
  $z\ge 3$ are the most massive galaxies with $E(B-V)\sim 0.15$ 
  at each epoch.   

\end{abstract}



\section{Introduction}
\label{section:intro}

Numerical simulations of galaxy formation evolve a comoving volume of
the Universe, starting from an initial state given by the theory of
inflation.  Such simulations are in principle capable of accurately
predicting the properties of galaxies that form from these initial
conditions, but limited computer resources impose severe restrictions
on the resolution and volume size that can be reached.  In general, it
is easier to achieve sufficient numerical resolution for high redshift
galaxies because the Universe is young and simulations are evolved
forward in time from the Big Bang.  However, observational surveys are
often mainly limited to low redshifts, looking outwards and backwards
in time from our vantage point.  In recent years, significant
advances in both observational and numerical techniques have
created an optimal overlap range between the two approaches at
intermediate redshifts ($z=2-6$), which is therefore a promising epoch
for comparing theoretical predictions with observations. 
This provides a testing ground for the current standard paradigm of
hierarchical galaxy formation in a universe dominated by cold dark
matter.

In observational surveys, one of the most important techniques for
detecting galaxies at redshifts $z \approx 3-6$ makes use of the Lyman
break, a feature at $\Ly = 4\pi \hbar^3 c / (m_e e^4) = 911.7634$
\AA\, (where $m_e$ is the reduced electron mass), the wavelength below
which the ground state of neutral hydrogen may be ionised. Blueward of
the Lyman break, a large amount of flux is absorbed by neutral
hydrogen, either in the galaxy itself or at some redshift along the
line of sight.  For the range we are interested in, the Lyman break is
redshifted into the optical part of the spectrum.  Because of this,
these galaxies can be detected using optical photometry, making them
attractive for ground-based surveys; a large difference between
magnitudes in nearby filters can give an estimate of the
observer-frame wavelength of the Lyman break, and thus the redshift of
the galaxy. A galaxy detected in this manner is called a Lyman Break
Galaxy (LBG).  The above `break' feature in a redshifted galaxy
spectrum causes it to fall in a particular location on the
colour-colour plane of, e.g., $U_n-G$ versus $G-R$ colour for $z
\approx 3$.  This colour-selection allows one to preselect the
candidates of high-redshift LBGs very efficiently
\citep[e.g.,][]{Steidel93, Steidel99}.

This method of detecting high-redshift galaxies has both advantages
and disadvantages. While it is capable of detecting a large number of
galaxies in a wide field of view using relatively little observation
time, it cannot assign exact redshifts to galaxies without follow-up
spectroscopy. Instead, it merely places LBGs into wide redshift bins.
Moreover, there is some concern that the procedure may introduce a
bias by preferentially selecting galaxies with prominent Lyman breaks.
These caveats should be kept in mind \citep[e.g.,][]{Ouchi04a} when
using the results from LBG observations for describing the general
characteristics of galaxies at high redshifts.  Nevertheless, the
efficiency of selecting high-redshift galaxy candidates coupled with
photometric redshift estimates can yield large samples that cannot be
obtained otherwise.  Using these techniques as well as follow-up 
spectroscopy, volume limited surveys of LBGs at $z\ga 3$ have been 
constructed \citep[e.g.][]{Lowenthal, Dickinson, Giavalisco}, some 
with a sample size of $N\sim 1000$ galaxies 
\citep[e.g.,][]{Steidel03, Ouchi04a}.

These large datasets make possible interesting comparisons with
numerical simulations.  Perhaps the most important fundamental
statistical quantity to consider for such a comparison is the
luminosity function of galaxies; i.e.~the distribution of the number
of galaxies with luminosity (or magnitude) per comoving volume.  We
will focus on this statistic here, as well as on the colours, stellar
masses, and number density of galaxies.

There have already been several previous studies of the properties of
LBGs using both semianalytic models \citep[e.g.,][]{Baugh,
Somerville, Blaizot}, and cosmological simulations
\citep[e.g.,][]{Nag02, Wei02, Harford03, NSHM}. 
As for the semianalytic models of galaxy formation, 
both \citet{Baugh} and \citet{Blaizot} were able to reproduce the 
number density and the correlation function of LBGs, and agree 
that LBGs at $z\sim 3$ are massive galaxies located in halos of 
mass $\sim 10^{12}\Msun$. But \citet{Somerville} emphasised that
merger induced starbursts 
(e.g. Mihos \& Hernquist 1996; Springel et al. 2005b) 
could also account for the observed number 
density of LBGs, therefore low mass galaxies with stellar masses
$\Mstar \sim 10^8\,\Msun$ could also contribute to the LBG 
population. 

Using numerical simulations, \citet{NSHM} studied the photometric
properties of simulated LBGs at $z=3$ including luminosity functions,
colour-colour and colour-magnitude diagrams using the same series of
SPH simulations described here \citep[see the erratum of the paper as
well:][]{NSHM-erratum}. They found that the simulated galaxies have
$U_n-G$ and $G-R$ colours consistent with observations (satisfying the
colour-selection criteria of Steidel et al.), when a moderate dust
extinction of $E(B-V)=0.15$ is assumed locally within the LBGs. In
addition, the observed properties of LBGs, including their number
density, colours and luminosity functions, can be explained if LBGs
are identified with the most massive galaxies at $z=3$, having typical
stellar masses of $\Mstar \sim 10^{10}\himsun$, a conclusion broadly
consistent with earlier studies based on hydrodynamic simulations of
the $\Lambda$CDM model. \citet{Nachos2, Nachos3} also extended 
the study down to $z=1-2$ using the same simulations, and focused on 
the properties of massive galaxies in UV and near-IR 
wavelengths.\footnote{After the submission of our manuscript, a preprint 
by \citet{Finlator} appeared on astro-ph, which studied the properties 
of simulated galaxies at $z=4$ in the same simulations used in this 
paper. Overall their results agree well with ours where comparison 
is possible. }

In this paper, we extend the work by \citet{NSHM} to the LBGs at even
higher redshifts $z=4-6$, focusing on the colours and luminosity
functions of galaxies. The paper is organised as follows.  In
Section~\ref{section:sims} we briefly describe the simulations used in
this paper, and in Section~\ref{section:method} we outline in detail
the methods used to derive the photometric properties of the simulated
galaxies.  We then present colour-colour diagrams in
Section~\ref{section:cc}, discuss stellar masses and number densities
of galaxies in Section~\ref{section:mass}, and present luminosity
functions in Section~\ref{section:lf}. Finally, we conclude in
Section~\ref{section:conclusion}.


\section{Simulations}
\label{section:sims}

The simulations analysed in this paper were performed
with the Smoothed Particle Hydrodynamics (SPH)
code Gadget-2 (Springel 2005), a Lagrangian 
approach for modeling hydrodynamic
flows using particles.  We employ the `entropy formulation'
of SPH \citep{SH02} which alleviates numerical overcooling problems
present in other formulations
(see e.g. Hernquist 1993; O'Shea et al. 2005). 
The simulation code allows for star
formation by converting gas into star particles on a characteristic
timescale determined by a subresolution model for the interstellar
medium \citep{SH03a}, which is invoked for sufficiently dense gas. In
this model, the energy from supernova explosions adds thermal energy
to the hot phase of the interstellar medium (ISM) and evaporates cold
clouds.  Galactic winds are introduced as an
extension to the model and provide a channel for transferring energy
and metal-enriched material out of the potential wells of galaxies
\citep[see][for more details]{SH03a}.  Finally, a uniform UV
background radiation field is present, with a modified
\citet{Haardt96} spectrum \citep{Dave1999, Katz1996}.  The simulations
are based on a standard concordance $\Lambda$CDM cosmology with
cosmological parameters $(\Omega_m, \Omega_\Lambda, \Omega_b,
\sigma_8, \hc) = (0.3, 0.7, 0.04, 0.9, 1)$, where $\hc = H_0/ (70
\kmsmpc)$.  We assume the same cosmological parameters when the
estimates of the effective survey volumes are needed for the
observations.

\begin{table*}
  \begin{center}
  \label{table:simulation}
  \begin{tabular}{ccccccccccc}  
    \hline
    Size & $L_{\rm box}$ & Run & $N_P$ & $m_{\rm DM}$ & $m_{\rm gas}$ & $\Delta \ell$ & $N(6)$ & $N(5)$ & $N(4)$ & $N(3)$ \\
    \hline
    Large & 142.9 & G6 & $486^3$ & $8.99\times 10^8$ & $1.38\times 10^8$ & $7.61$ & 12570 & 27812 & 48944 & 74414 \\
    Medium&  48.2 & D5 & $324^3$ & $1.16\times 10^8$ & $1.80\times 10^7$ & $5.96$ & 6767 & 11991 & 18897 & 24335 \\
    Small &  14.3 & Q6 & $486^3$ & $8.99\times 10^5$ & $1.38\times 10^5$ & $1.17$ & 9806 & 12535 & 15354 & \\
          &       & Q5 & $324^3$ & $3.03\times 10^6$ & $4.66\times 10^5$ & $1.76$ & & & & 7747 \\
    \hline
  \end{tabular}
  \medskip
  \caption{
  Simulation parameters and number of galaxies identified. The
  simulation boxsize $L_{\rm box}$ is given in $\himsun$. The (initial)
  number of gas particles $N_P$ is equal to the number of dark matter
  particles, so the total particle count is twice $N_P$. The mass of each
  particle ($m_{DM}$ for dark matter and $m_{gas}$ for gas) is given in
  $\himsun$.  The softening length $\Delta \ell$ is given in ${\rm
  \hinv kpc}$.  The last four columns give the number of galaxies, $N(z)$,
  found in the simulation at a given redshift $z$ by the group finder. For the
  `Small' boxsize, simulation Q6 was used for $z\geq 4$, while Q5 was used for
  $z=3$.}
  \end{center}
\end{table*}

For this paper, we analyse the outputs of simulations with three
different box sizes and mass resolutions at redshifts $z=3-6$.  The
three simulations employed belong to the G-series, D-series, and
Q-series described in \cite{SH03b}, with corresponding box sizes of
142.9, 48.2, and 14.3 $\himpc$ in comoving coordinates. We shall refer
to them as `Large', `Medium', and `Small' simulations, respectively.
The primary differences between the three runs are the size of the
simulation box, the number of particles in the box, and hence the mass
of the individual particles (i.e. the mass resolution). The parameters
for each run are summarised in Table~1. The same simulations were used
for the study of the cosmic star formation history \citep{SH03b,
Nachos1}, LBGs at $z=3$ \citep{NSHM}, damped Lyman-$\alpha$ systems
\citep*{NSH04a, NSH04b}, massive galaxies at $z=2$ \citep{Nachos2,
Nachos3}, and the intergalactic medium
\citep{Fetal04a, Fetal04b, Fetal04c,
Fetal04d}.

Having three different simulation volumes allows us to assess the
effect of the boxsize on our results. The measured luminosity
functions suffer from two different types of resolution effects. On
the bright end, the boxsize can severely limit a proper sampling of
rare objects, while on on the faint end, the finite mass resolution
can prevent faint galaxies from being modelled accurately, or such
galaxies may even be missed entirely.  We will discuss these two
points more more explicitly in Section 5.


\section{Method}
\label{section:method}

Galaxies were extracted from the simulation by means of a group finder
as described in \cite{NSHM}. The group finder works by first smoothing
the gas and stellar particles to determine the baryonic density
field. The particles which pass a certain density threshold for star
formation are then linked to their nearest neighbour with a higher
density, unless none of the 32 closest particles has a higher
density. This is similar to a friends-of-friends algorithm, except it
does not make use of a fixed linking length.

\begin{figure}
\includegraphics[width=84mm]{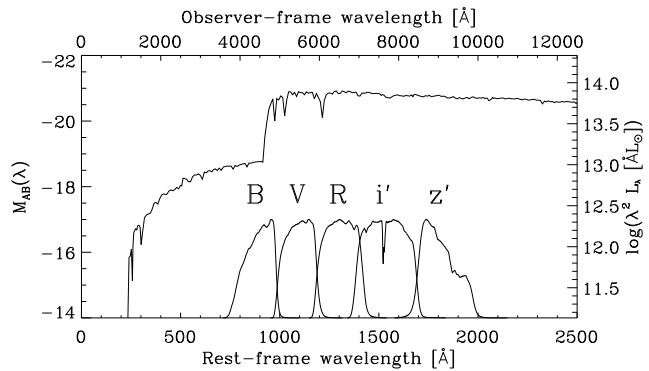}
\caption{Sample LBG spectrum, taken from the `Medium'-size simulation 
  at $z=4$. The bottom axis gives the intrinsic wavelength of the spectrum,
  and the top axis is the wavelength redshifted by a factor of $1+z$. Also
  shown are the response functions of five Subaru filters (Johnson $B$, $V$,
  and $R$, and SDSS $i'$ and $z'$), positioned at their effective wavelengths.
}
\label{Spectrum}
\end{figure}

After the galaxies are identified, several steps are taken to derive
their photometric properties. First, we compute the spectrum of
constituent star particles based on their total mass and metallicity,
using a modern population synthesis model of \citet{BC2003}. The
spectral energy distribution (SED) is given as a set of ordered pairs
$(\lambda, L_\lambda(\lambda))$. The sampling resolution of this
function varies, but in the region of interest it is approximately
10\AA. A sample spectrum from the `Medium'-size simulation (D5) at
$z=4$ is shown in Fig.~\ref{Spectrum}.

Once the intrinsic SED for each source is computed, it must be
transformed to represent the spectrum as it would be seen by an
observer on Earth. This process involves several steps. First, we
apply the Calzetti dust extinction law to the spectrum
\citep{Calzetti}, which accounts for intrinsic extinction within the
galaxy. The specific values we adopt for the strength of the
extinction will be discussed in more detail below.  Next, we redshift
the spectrum and account for IGM absorption \citep{Madau}.  Finally,
we compute the photometric magnitudes by convolving the resulting SED
with various filter functions. This allows us to determine the
apparent magnitude of each object for commonly employed filters in the
real observations.

The formula used for this computation may be derived as follows.  For a given
source (defined by its flux per unit frequency $f_\nu(\nu)$), observed through
a given filter (defined by its filter response function $R(\nu)$), the
apparent magnitude is given by \citep[][Eq. 7]{Fukugita}:
\begin{equation}
m = -2.5\log{\int f_\nu (\nu) R(\nu)\, {\rm d}\ln \nu \over \int C_\nu (\nu)
  R(\nu)\, {\rm d} \ln \nu},
\end{equation}
where $C_\nu(\nu)$ is the reference SED. For the AB magnitude system,
$C_\nu$ is a constant ($10^{-19.44}$ erg s$^{-1}$ cm$^{-2}$
Hz$^{-1}$), and for the Vega system, $C_\nu$ is the SED of the star
Vega.   The above formula may be rewritten in terms of
wavelength using the relation $f_\nu(\nu) = (\lambda^2 /
c) f_\lambda(\lambda)$, where $c$ is the speed of light,
and the observed flux may be related to the intrinsic luminosity by
$f_\lambda(\lambda) = L_\lambda(\lambda / (1+z)) / [4\pi d_L^2
(1+z)]$, where $\lambda$ is the wavelength in the observer frame, and
$d_L$ is the luminosity distance to redshift $z$. This gives (for cgs
units):
\begin{equation}
m_{AB} = -2.5\log{\int \lambda L_\lambda\, ({\lambda \over 1+z})\,
  R(\lambda)\, {\rm d}\lambda \over 4\pi\, d_L^2\, c\, (1+z) \int {1\over \lambda}
  R(\lambda)\, {\rm d}\lambda} - 48.60, 
\end{equation}
The absolute magnitude $M_{AB}$ may be determined from this equation by
setting $z=0$ and $d_L = 10$ pc. For monochromatic magnitudes, as shown in
Fig.~\ref{Spectrum}, the equation reduces to $M_{AB}(\lambda) = -2.5
\log(\lambda^2 L_\lambda) + 13.83$, assuming $\lambda$ is given in units of
[\AA] and $L_\lambda$ is in units of [${\rm L_{\odot}\, \AA}^{-1}$].

Absorption by dust and extinction by the IGM each add a multiplicative
factor to $f_\lambda$ as a function of wavelength inside the
integral. For dust absorption, the factor is
$10^{-k(\lambda/(1+z))E}$, where $k(\lambda/(1+z))$ is the Calzetti
extinction function, and $E \equiv E(B-V)$ is the extinction in $B-V$
colour, taken to be a free parameter.  There is no simple theoretical
constraint on $E(B-V)$ except that it must be nonnegative, so we
simply consider a range of values for $E(B-V)$ to study the extinction
effect systematically.  Since the latest surveys
\citep[e.g.][]{Shapley} suggest that $E(B-V)$ ranges from 0.0 to
0.3 with a mean of $\sim 0.15$, we adopt three fiducial values of
$E(B-V)=0.0$, 0.15, and 0.30. In most of our figures, they
will be indicated by the colours blue, green, and red,
respectively. We discuss a different choice for the assignment of
extinction to galaxies in Section~\ref{section:lf}.

\begin{figure}
\begin{center}
\includegraphics[width=84mm]{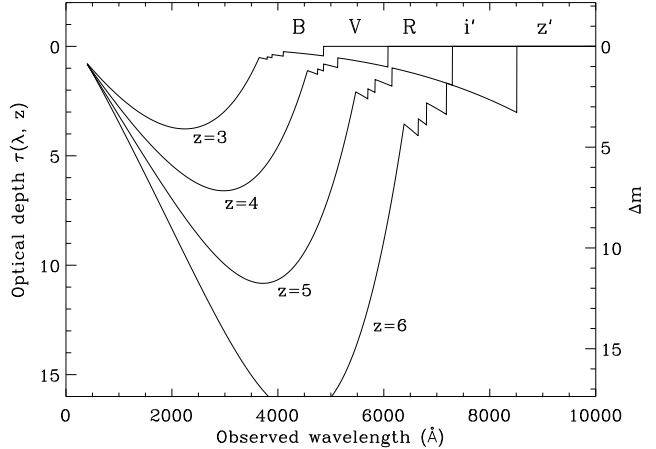}
\end{center}
\caption{Total optical depth owing to IGM absorption, at 
  redshifts of 6, 5, 4, and 3, taking into account continuum absorption and
  absorption from the four strongest lines. The right axis gives the
  corresponding increase in apparent magnitude. The names of the five Subaru
  filters are positioned at their effective wavelengths.}
\label{Madau}
\end{figure}

For the IGM extinction, the factor is $\exp[-\tau(\lambda, z)]$, 
where $\tau(\lambda, z)$ is the effective optical depth owing to both 
continuum \citep[][ footnote 3]{Madau} and line extinction 
\citep[][ Eq. 15]{Madau}:
\begin{eqnarray}
\tau(\lambda, z) & = & 0.25x_c^{3.46}(a^{0.46}-1) \nonumber \\
 & + & x_c^{1.68}(9.4a^{0.18}+0.7a^{-1.32}-0.023a^{1.68}-10.077) \nonumber \\
 & + & \sum_{j=2}A_{j}(x_c{j^2-1\over j^2})^{3.46},
\label{eq:tau}
\end{eqnarray}
where $x_c \equiv \lambda/\Ly$, $a \equiv (1+z)/x_c$, and the $A_j$ 
are the line strength coefficients. We consider only the four 
strongest lines, corresponding to $j=2$ to 5. This absorption 
becomes highly significant in the blue bands at redshifts greater 
than 3, as shown in Fig.~\ref{Madau}.

Note that dust absorption is applied in the rest frame of the galaxy,
while IGM extinction is applied in the observer's frame.  The integration
can also be done in the rest frame rather than the observer frame by
substituting $\lambda$ with $\lambda (1+z)$.  Thus the overall formula to
compute $m$ from $L_\lambda$ is:
\begin{eqnarray}
m_{AB} = -48.60 - 2.5\log \Bigl({1+z \over 4\pi d_L^2\, c} \hspace{30mm}
 \nonumber \\
\hspace{8mm}
{\int \lambda L_\lambda(\lambda)\, 10^{-k(\lambda)E}\,
  e^{-\tau(\lambda(1+z), z)}\, R(\lambda (1+z))\, {\rm d}\lambda \over
\int {1 \over \lambda}\, R(\lambda (1+z))\, {\rm d}\lambda } \Bigr)
\end{eqnarray}

We have used several filters for our calculations, each defined by a
response function $R(\lambda)$. Specifically, for comparison with
observations from the Subaru telescope, we used their filters $B$,
$V$, $R$, $i'$, and $z'$ \citep[Johnson-Morgan-Cousins system; see 
Section 2.6 of][]{Miyazaki}, which provide good coverage of all optical
wavelengths, and some into the near-infrared. 
\citet{Ouchi04a} treated the $i'$ magnitude as the standard UV
magnitude for $z \sim 4$, and the $z'$ magnitude as the standard UV
magnitude for $z \sim 5$.

For comparison with surveys that did not use the Subaru filters, and 
for a more general UV luminosity function, we used a boxcar-shaped filter
(i.e. response function set equal to unity) centered at 1700\AA\, and
with a half-width of 300\AA, in the rest frame of the observed
galaxy. Note that this is actually a different filter in the
observer's frame depending on the redshift of the observed object
(e.g., for a $z=4$ object, it is centered at rest frame 8500\AA, and
for a $z=5$ object, it is centered at 10200\AA).  We refer to the
magnitude measured with this filter simply as the `UV-magnitude' in
this paper. Depending on a particular survey's capabilities,
observationally determined UV magnitudes may be based on a slightly
different wavelength than 1700\AA, but it is a reasonable assumption
that the resulting magnitudes are comparable.


\section{Colour-Colour Diagrams}
\label{section:cc}

\begin{figure*}
\includegraphics[width=80mm]{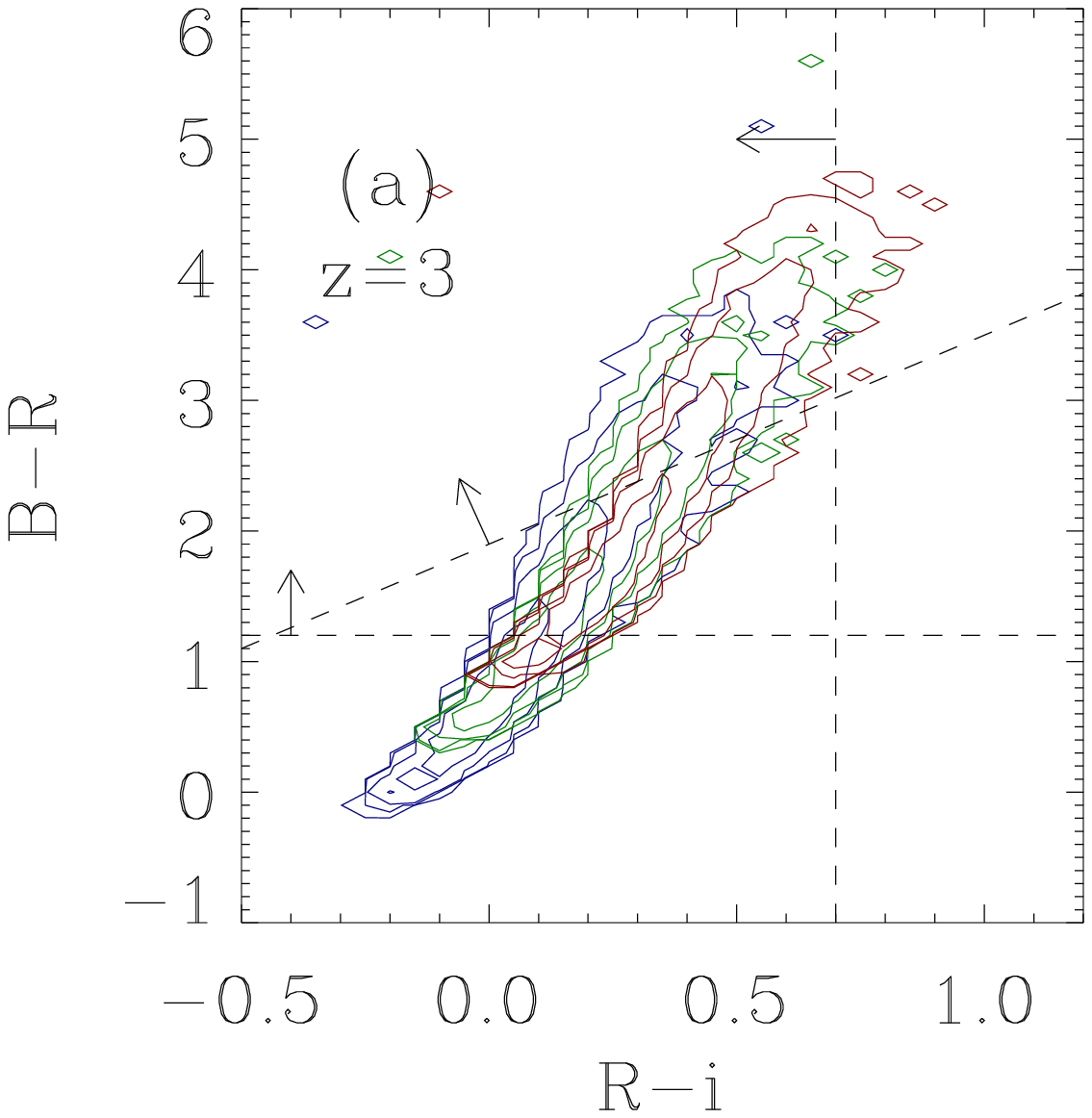}
\includegraphics[width=80mm]{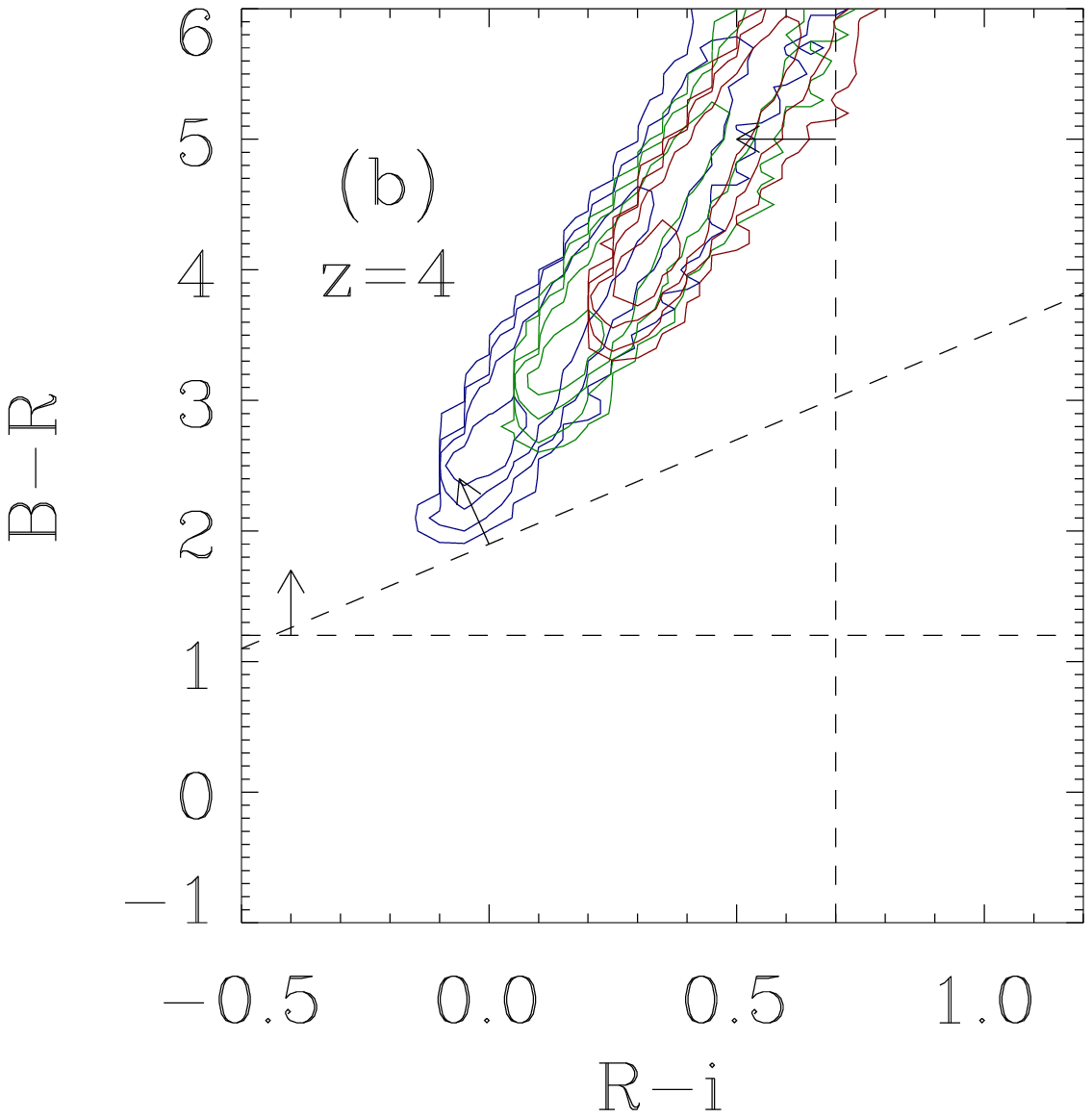}
\caption{Colour-colour diagram for the plane of $B-R$ versus $R-i'$.
  The selection criteria for $BRi$--LBGs in the Subaru survey are shown by the
  dashed lines. The contours in panels (a) and (b) show simulated galaxies in
  the `Large' (G6) run for $z=3$ and $z=4$, respectively. Blue, green, and red
  contours correspond to $E(B-V)=0.0$, 0.15, and 0.30, respectively.  }
\label{CC-BRi}
\end{figure*}

\begin{figure*}
\includegraphics[width=80mm]{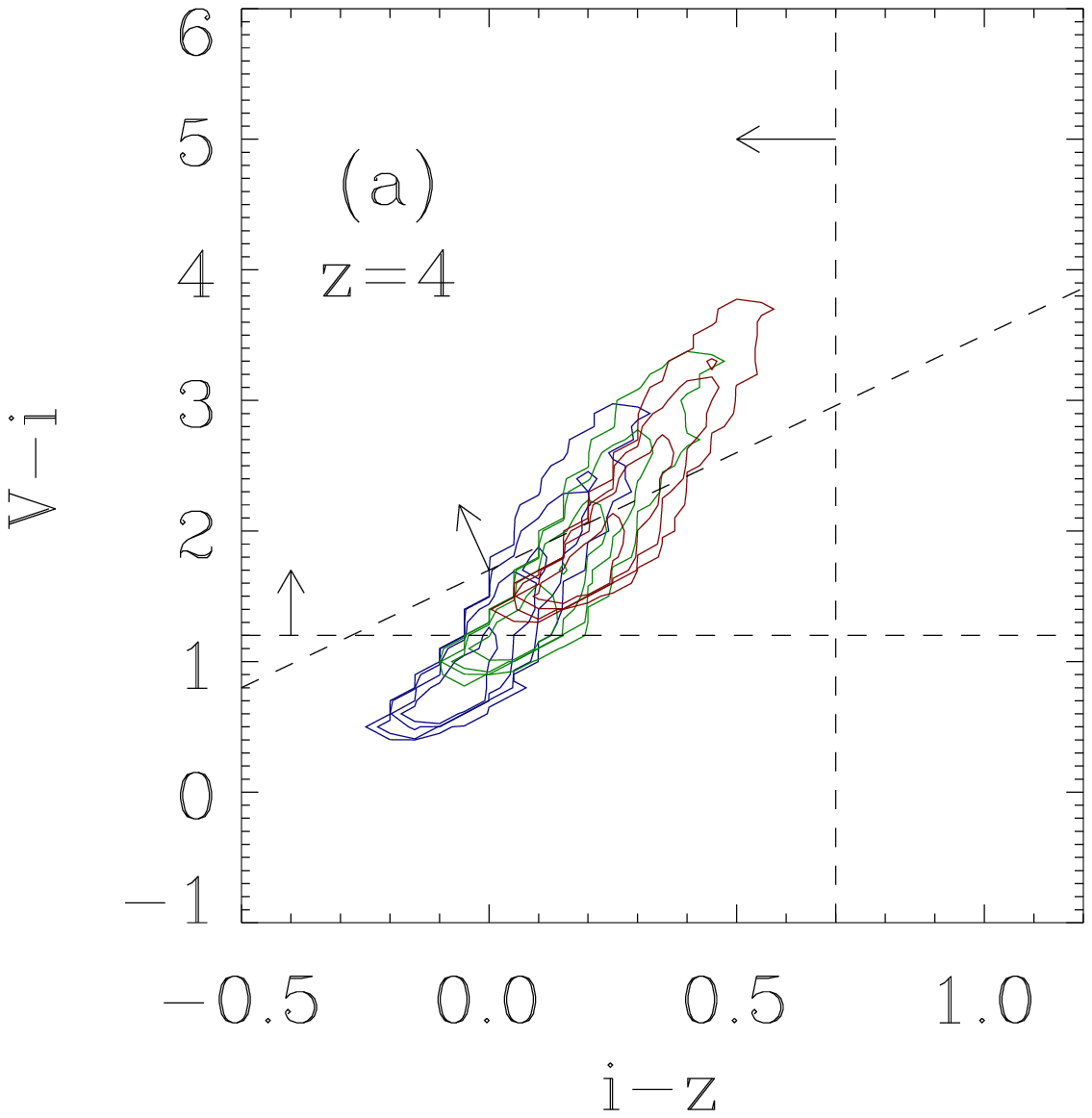}
\includegraphics[width=80mm]{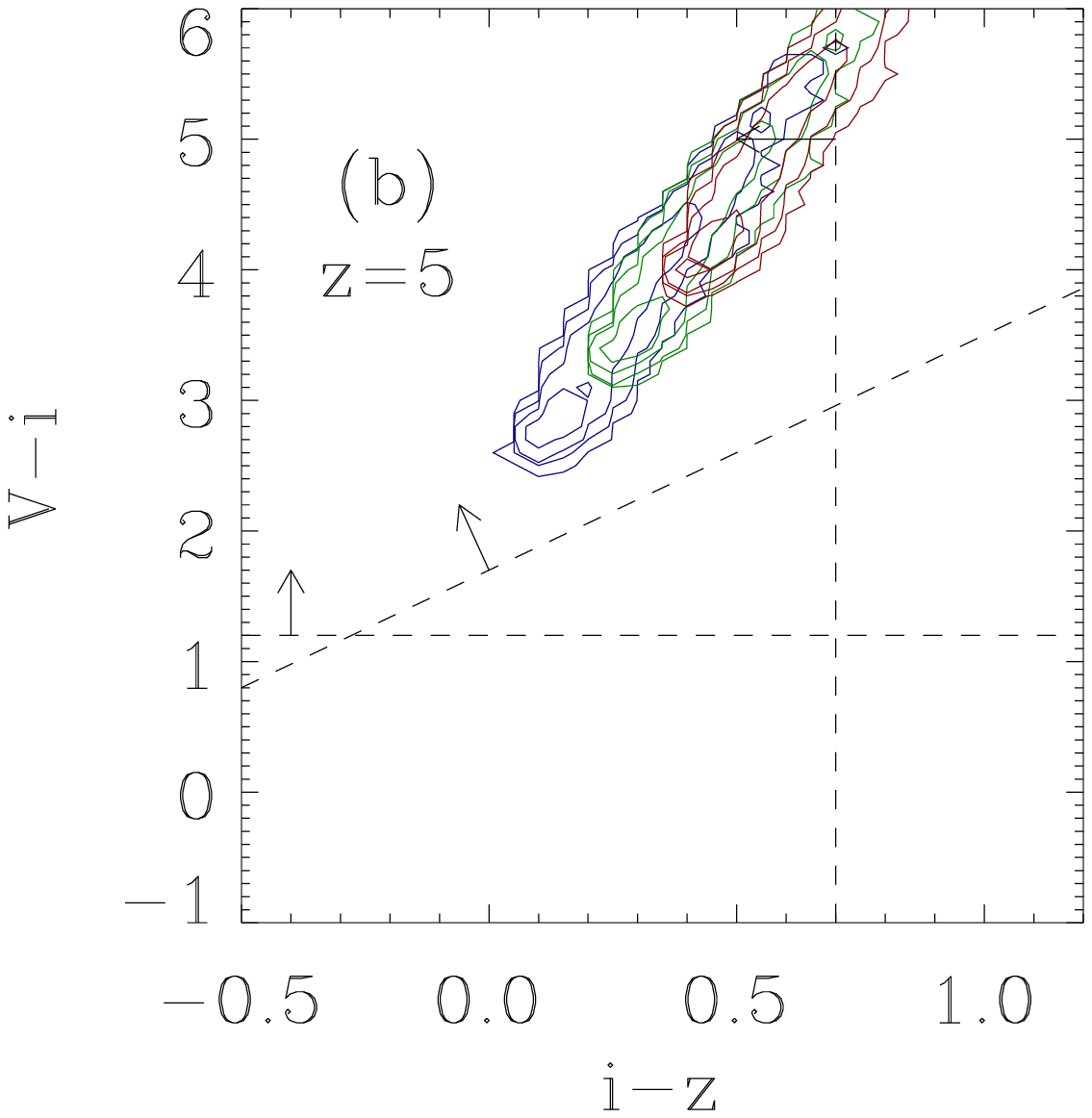}
\caption{Colour-colour diagram for the plane of $V-i'$ versus $i'-z'$. 
  The selection criteria for $Viz$--LBGs in the Subaru survey are shown by the
  dashed lines. The contours in panels (a) and (b) show the simulated galaxies
  in the `Large' (G6) run for $z=4$ and $z=5$, respectively. Blue, green, and
  red contours correspond to $E(B-V)=0.0$, 0.15, and 0.30, respectively.  }
\label{CC-Viz}
\end{figure*}

LBGs are identified based on a significantly dimmer magnitude in a
filter blueward of their Lyman break compared with a filter redward of
their Lyman break. This difference is manifest as a significantly
redder colour.  Moreover, two filters redward of the Lyman break
should not exhibit abnormal dropouts with respect to one another, a
fact that can distinguish them from interlopers with very red spectra.
Thus, in order to select LBGs from the sample, colour selection
criteria are very important. For instance, galaxies at $z\approx 4$
will have Lyman breaks at approximately 4600\AA, between the $B$ and
$R$ filters (as shown in Fig.~\ref{Spectrum}), so these galaxies
will have large $B-R$ colours.  But, we also expect them to have
moderate $R-i'$ colours, since both $R$ and $i'$ are redward of
4600\AA.  The exact colour-colour selection criteria used by each
survey are determined empirically by placing a sample of
spectroscopically identified LBGs on a colour-colour diagram, or
computing the track of local galaxies with known spectra (or
artificial spectra of galaxies generated by a population synthesis
model with assumed star formation histories) as a function of
redshift.

\begin{figure*}
\includegraphics[width=80mm]{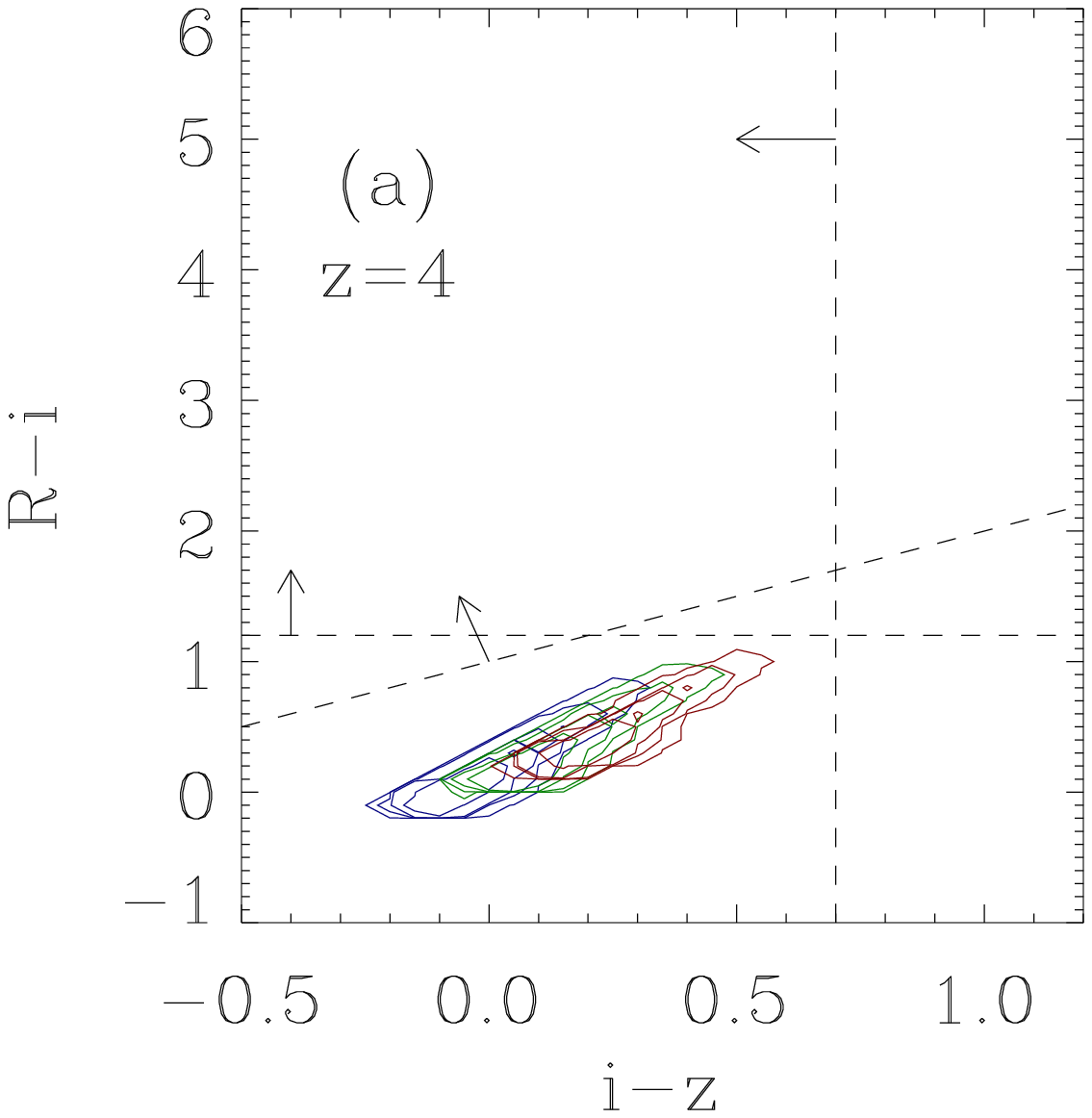}
\includegraphics[width=80mm]{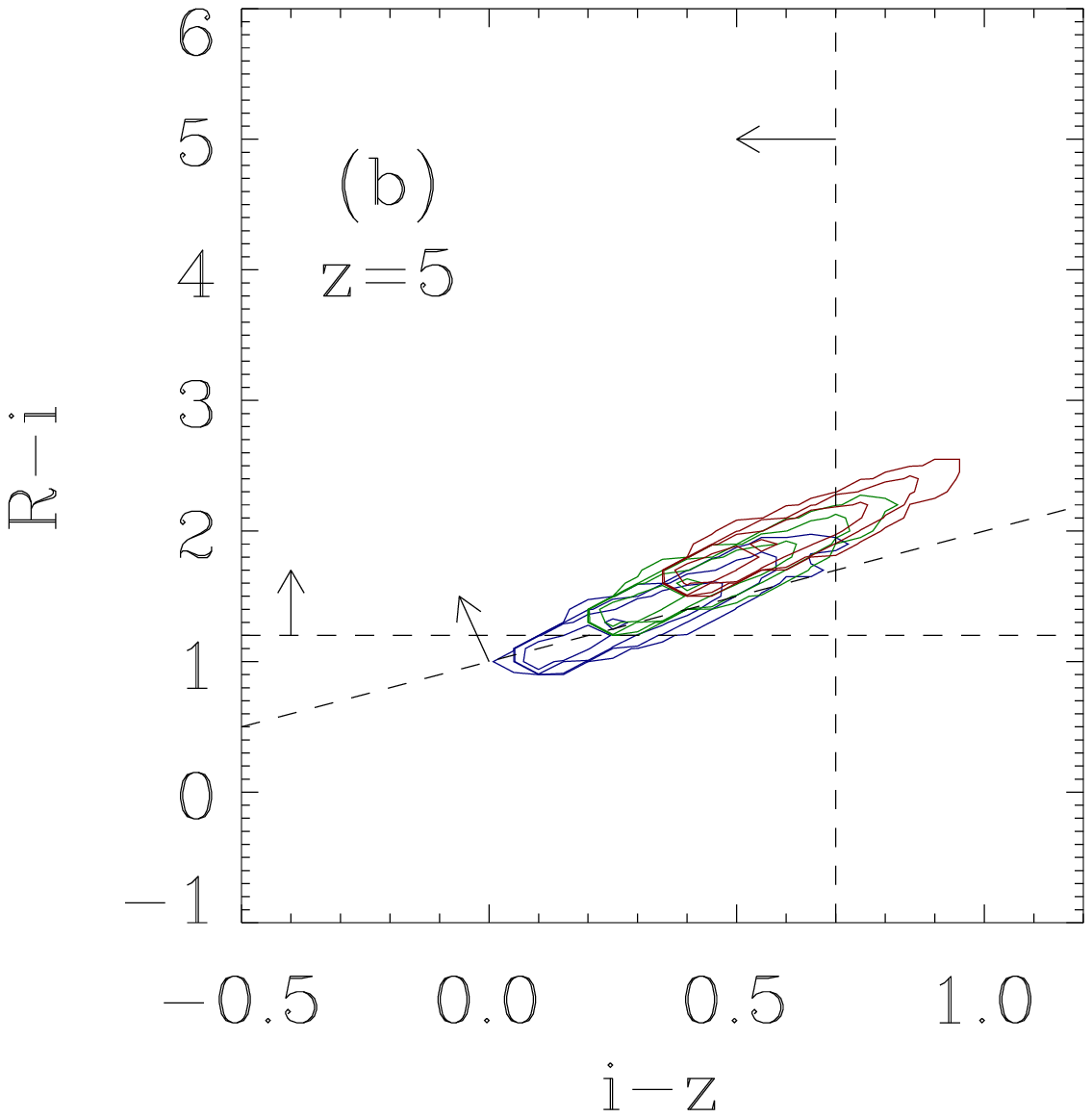}
\caption{Colour-colour diagram for the plane of $R-i'$ versus $i'-z'$. 
  The selection criteria for $Riz$--LBGs in the Subaru survey are shown by the
  dashed lines. The contours in panel (a) and (b) show the simulated galaxies
  in the `Large' (G6) run for $z=4$ and $z=5$, respectively. Blue, green, and
  red contours correspond to $E(B-V)=0.0$, 0.15, and 0.30, respectively.  }
\label{CC-Riz}
\end{figure*}

\begin{figure}
\begin{center}
\includegraphics[width=84mm]{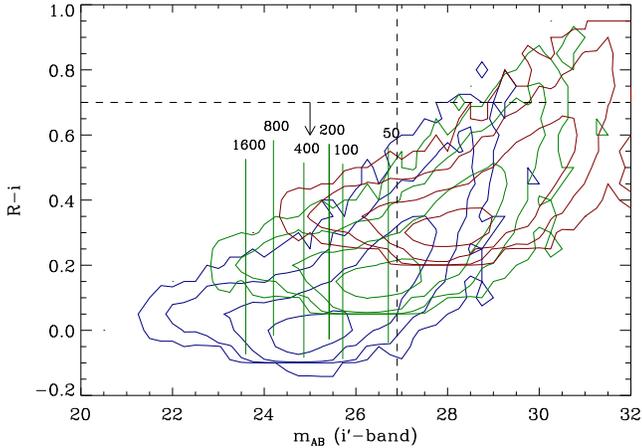}
\end{center}
\caption{Colour-magnitude diagram for the `Medium' boxsize (D5) run 
  at $z=4$. Blue, green, and red contours outline the distribution of galaxies
  for $E(B-V) = 0.0$, 0.15, and 0.30, respectively. Also shown in green
  are vertical lines approximately delineating the number of
  star particles in
  galaxies. Most galaxies found to the left of the first line comprise
  1600 star particles or more, and 200 particles mark the end of our
  interval of confidence. Black dashed lines are cutoffs for $BRi$--LBGs as
  observed by the Subaru group. The vertical line indicates the $3\sigma$
  limiting magnitude of $i'=26.9$, and the horizontal line represents one of
  the colour-colour selection criteria used to identify LBGs. }
\label{CMD-Dz4}
\end{figure}

For example, \citet{Ouchi04a} use the following colour
criteria for selecting $z\sim 4$ galaxies:
\begin{eqnarray}
\hspace{20mm} B-R & > & 1.2 \\
\hspace{20mm} R-i' & < & 0.7 \\
\hspace{20mm} B-R & > & 1.6(R-i')+1.9
\end{eqnarray}
Galaxies identified as LBGs by these criteria are called $BRi$--LBGs.
Similar criteria exist for $V-i$' versus $i'-z'$ colours, and $R-i'$ versus
$i'-z'$ colours; galaxies selected in this way are called $Viz$--LBGs and
$Riz$--LBGs, respectively. Each of these three classes of LBGs corresponds to
an approximate range in redshift space. The central redshifts for $BRi$,
$Viz$, and $Riz$ populations are approximately 4, 5, and 5, respectively. The
$Riz$ selection has a narrower range of redshift than the $Viz$ selection, so
we will use the $BRi$ sample to compare to our $z=4$ simulations, and $Riz$
sample to compare to our $z=5$ simulations.

In Figs.~\ref{CC-BRi} -- \ref{CC-Riz}, we show the colour-colour
diagrams of our simulated galaxies. Overall, the agreement with the
observation is good, and the simulated galaxies fall within the same
region as the observed galaxies, a conclusion consistent with
\citet{NSHM}. Very few $z=3$ galaxies in our simulation would be
detected as $BRi$--LBGs, but a large fraction of $z=4$ galaxies
would. Similarly, relatively few $z=4$ galaxies in our simulations
would be detected as $Viz$--LBGs or $Riz$--LBGs compared with $z=5$
galaxies. This result appears to be relatively insensitive to the
amount of Calzetti extinction, at least for the range of extinction
values we considered.

In Fig.~\ref{CMD-Dz4}, we also show the colour--magnitude diagram on
the plane of $i'$-band apparent magnitude and $R-i'$ colour for the
`Medium' (D5) run. This figure shows that all the simulated galaxies
brighter than $m_{AB}(i'-{\rm band})=27$ satisfy the colour-selection
of $R-i' < 0.7$. Since the brightest galaxies in the simulations are
the most massive ones, this means that the LBGs in the simulations are
the brightest and most massive galaxies with $E(B-V)\sim 0.15$ at each
epoch.  The situation of course changes when a larger value of
extinction is allowed, as such galaxies could become redder than
$R-i'=0.7$.  Such dusty starburst galaxies may exist in the real
universe, but we are not considering them in this paper by restricting
ourselves to $E(B-V)\leq 0.3$.


\section{Galaxy stellar masses}
\label{section:mass}

\begin{figure*}
\begin{center}
\includegraphics[width=80mm]{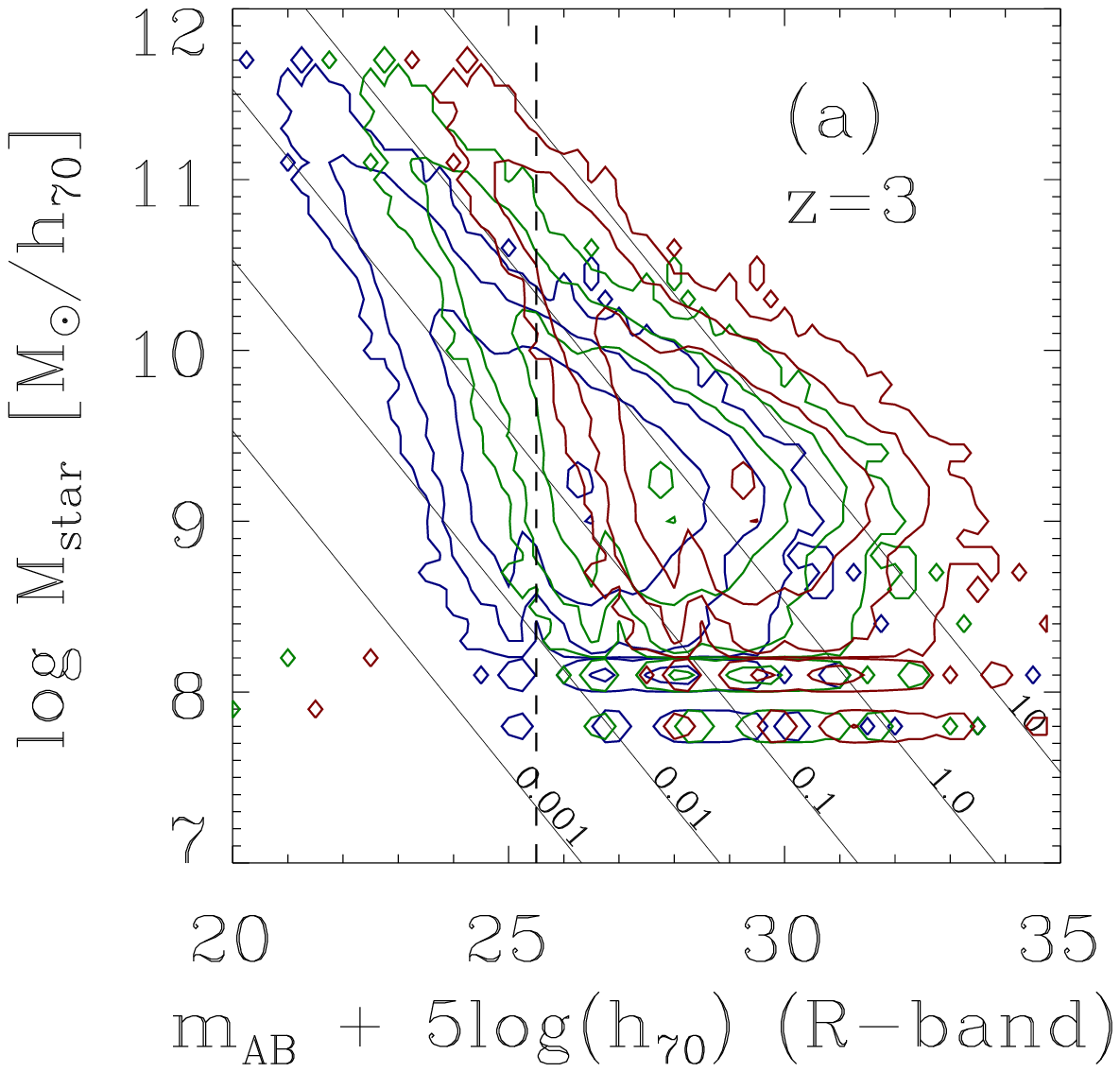}
\includegraphics[width=80mm]{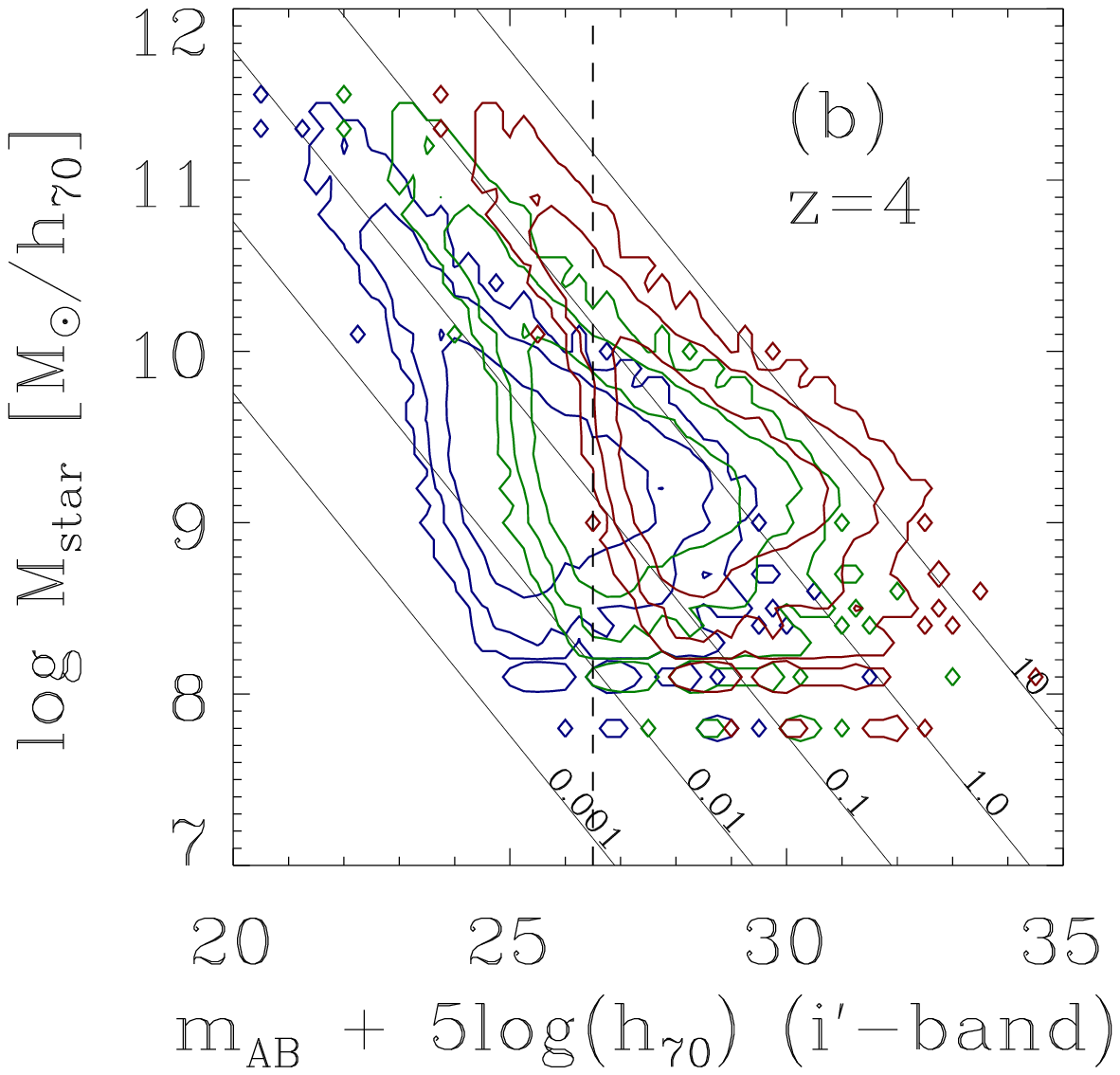}
\vspace{4mm} \\
\includegraphics[width=80mm]{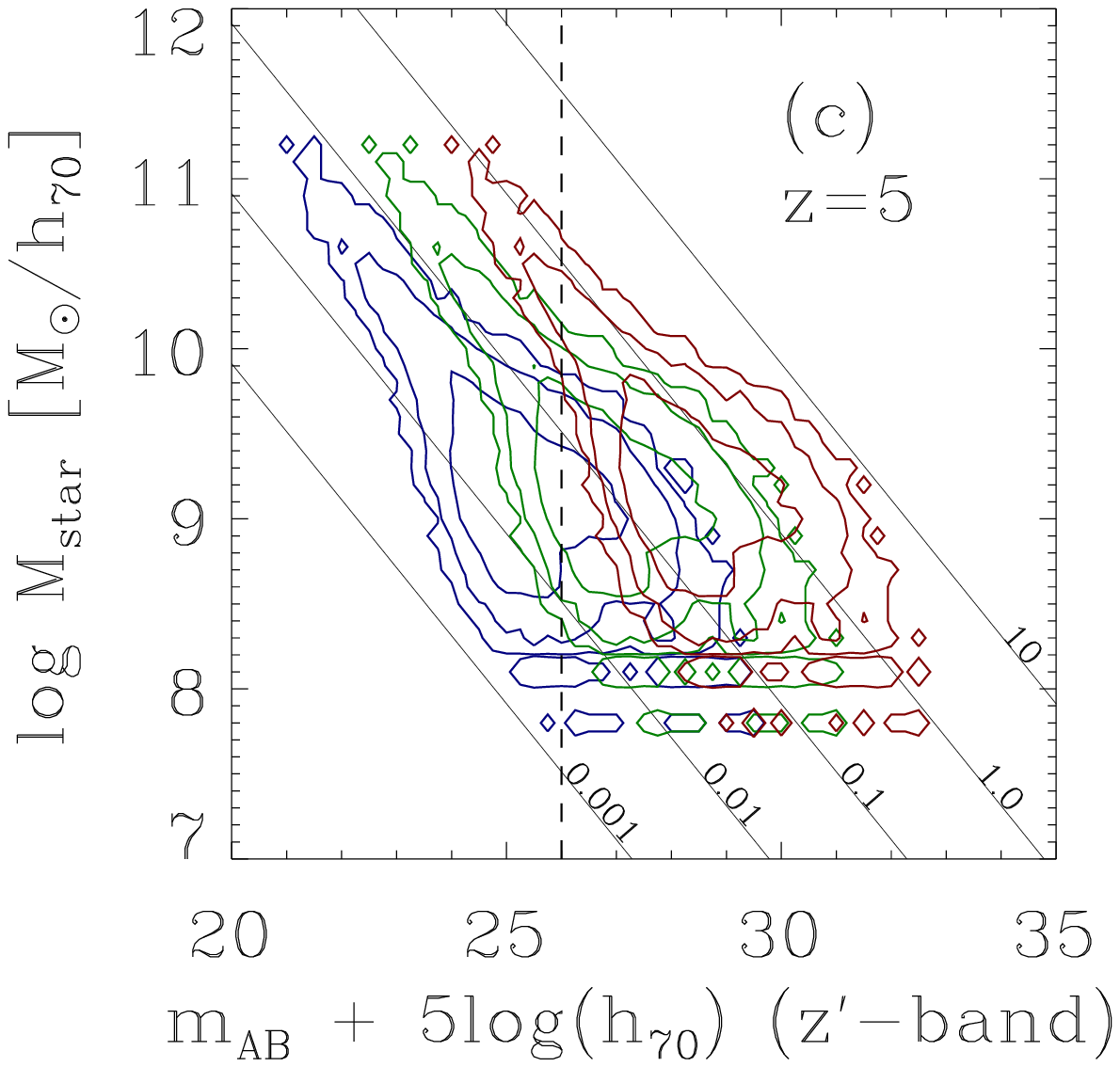}
\includegraphics[width=80mm]{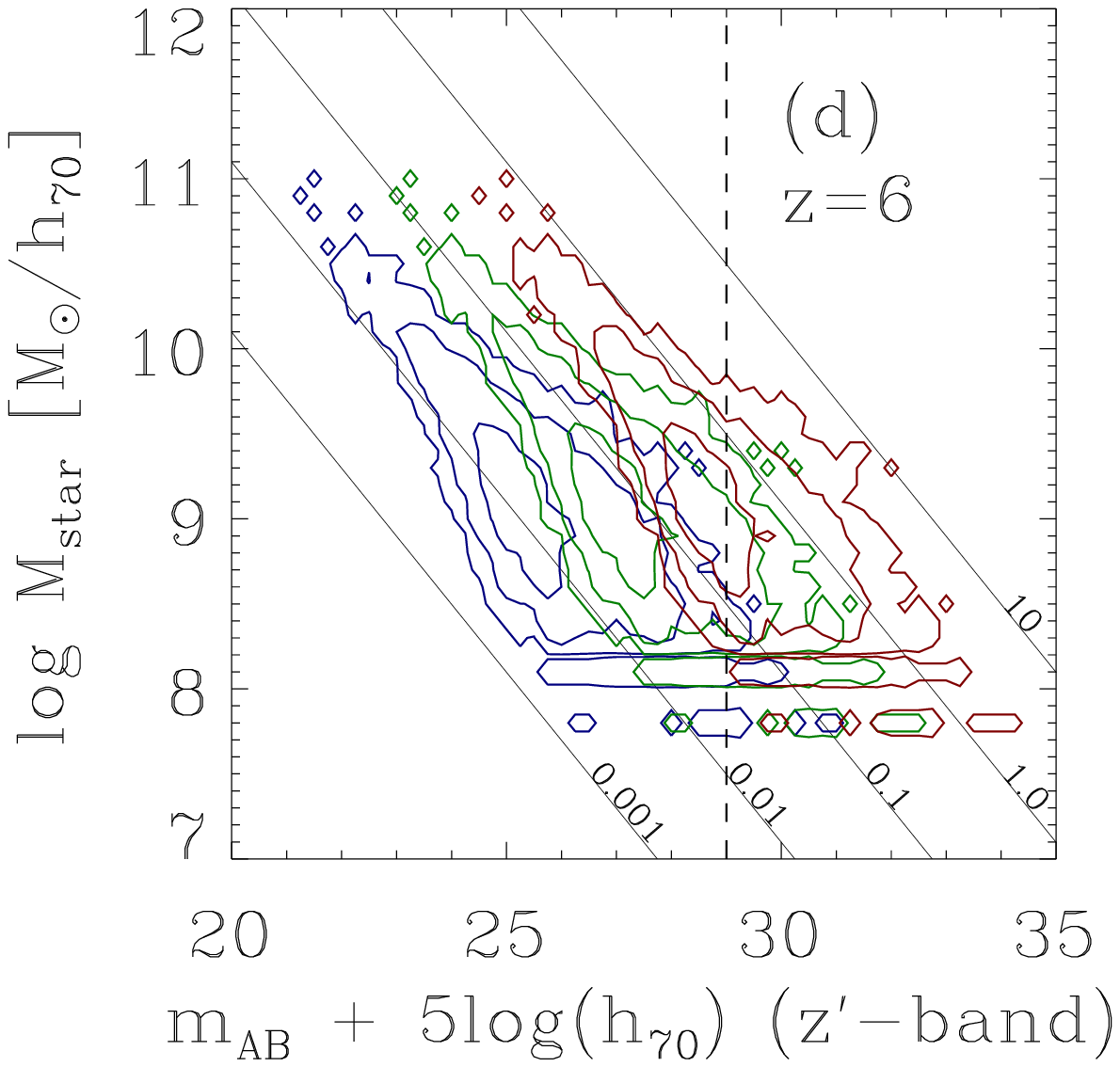}
\end{center}
\caption{
  Contour plots of stellar mass of galaxies vs. UV magnitude, for redshifts
  $z=3-6$, using the `Large' (solid colour contour) box size. Blue, green, and red
  contours represent extinctions of $E(B-V)=0.0$, 0.15, and 0.3 respectively.
  The dashed black lines indicate the magnitude limits of \citet{Steidel03}
  at $z=3$, the SDF sample for $z=4$ and 5, and the HST GOODS for $z=6$.
  Diagonal lines show lines of constant stellar mass to light ratio; the
  value of $\Mstar / (\lambda L_\lambda)$ is labeled for each line, in units
  of $\Msun / \Lsun$.  }
\label{Mass-scale}
\end{figure*}

Fig.~\ref{Mass-scale} shows stellar mass $\Mstar$ vs. UV magnitude
for the simulated galaxies over several redshifts and for different
extinction values, in relation to the survey limiting magnitudes. From
the figure it is seen that the `Large' box size simulation contains
galaxies as massive as $\Mstar \sim 10^{11}\himsun$ at $z=6$, and
$\Mstar \sim 10^{11.7}\himsun$ at $z=3$. Larger objects are too rare
to be found in a simulation of this size. The diagonal lines in the figure
depicting mass-to-light ratio show that this value is generally increasing
going from higher to lower redshift, so that the luminosity per stellar
mass is decreasing with time.

\begin{figure*}
\begin{center}
\includegraphics[width=80mm]{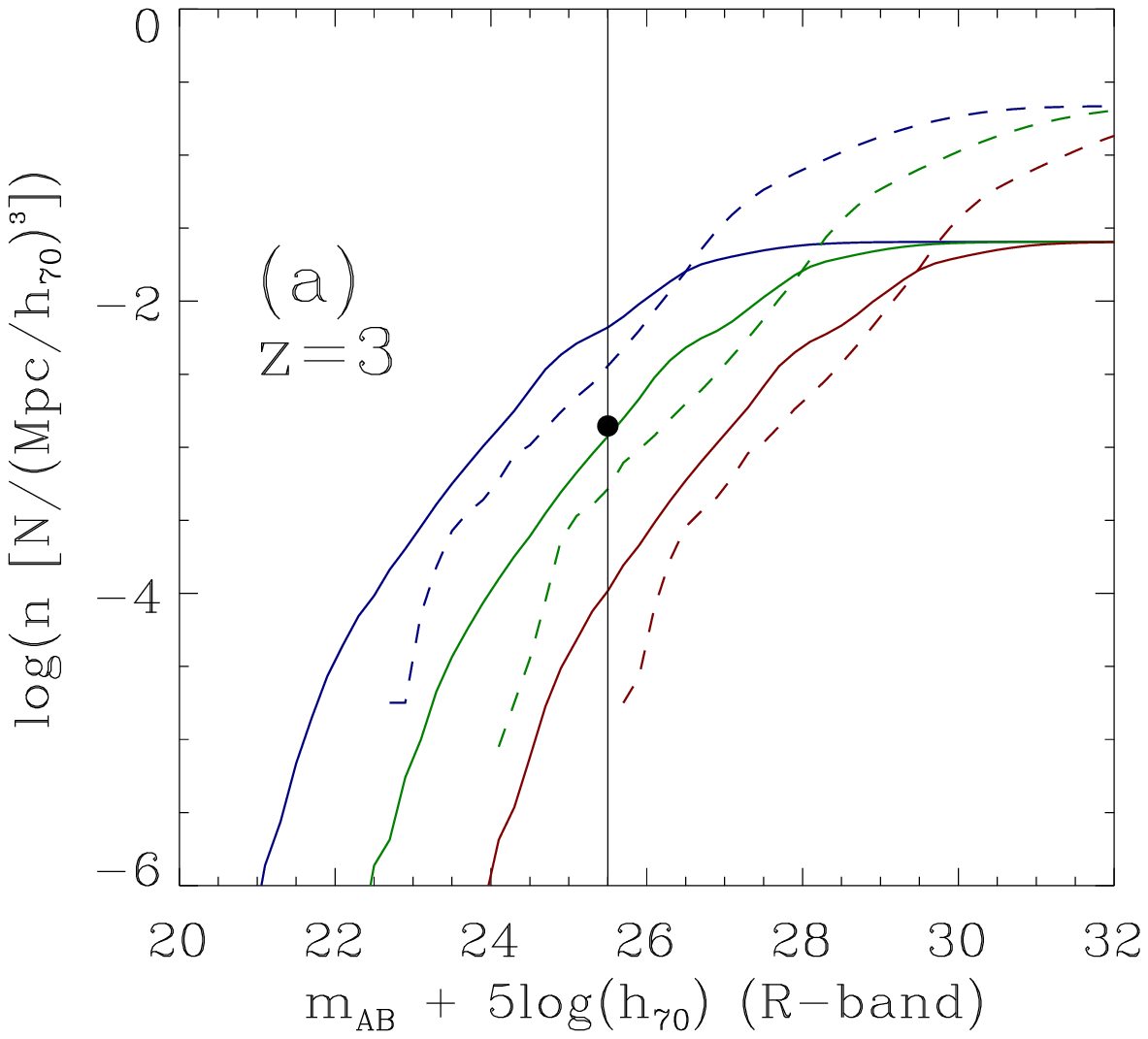}
\includegraphics[width=80mm]{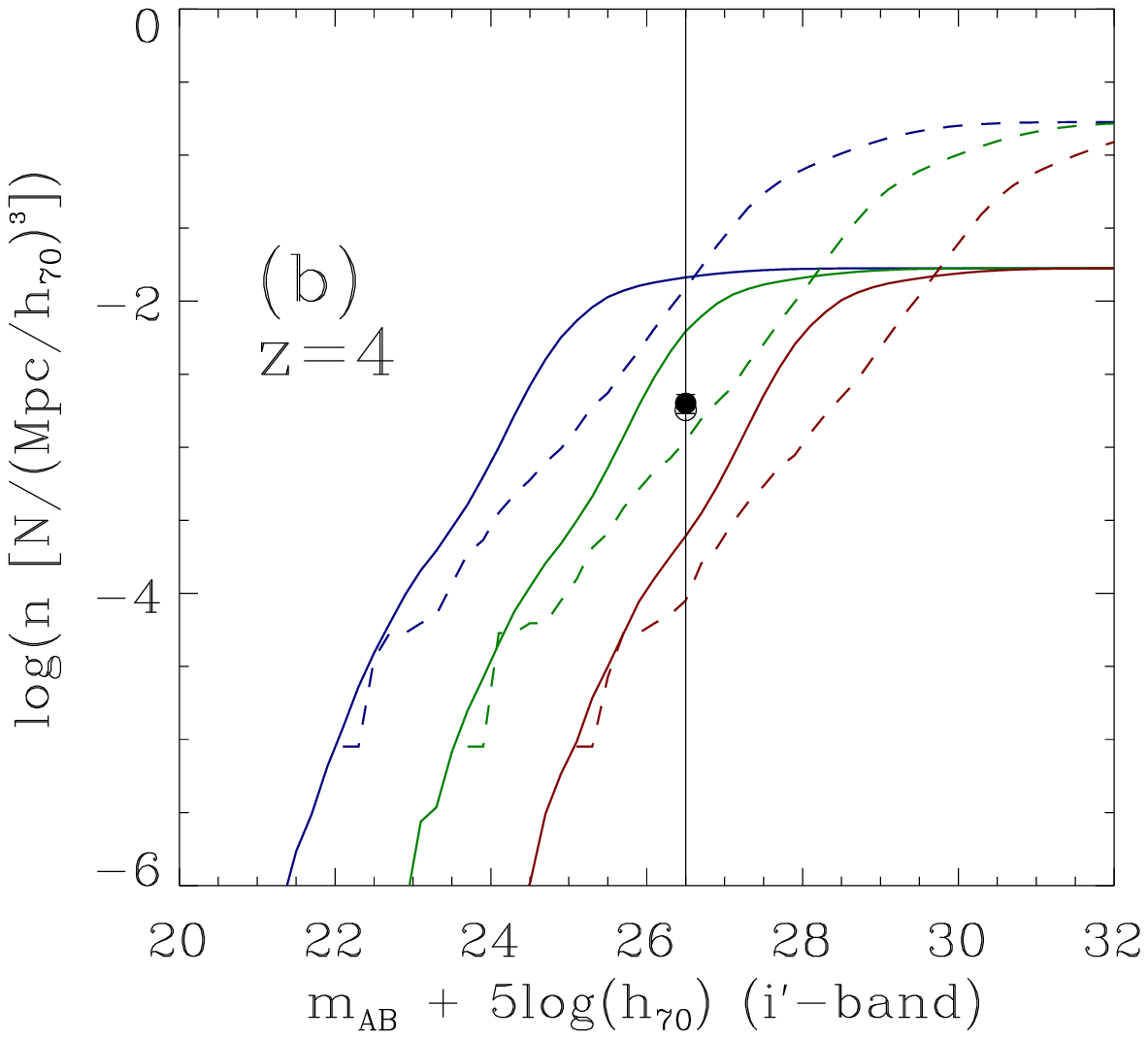}
\vspace{4mm} \\
\includegraphics[width=80mm]{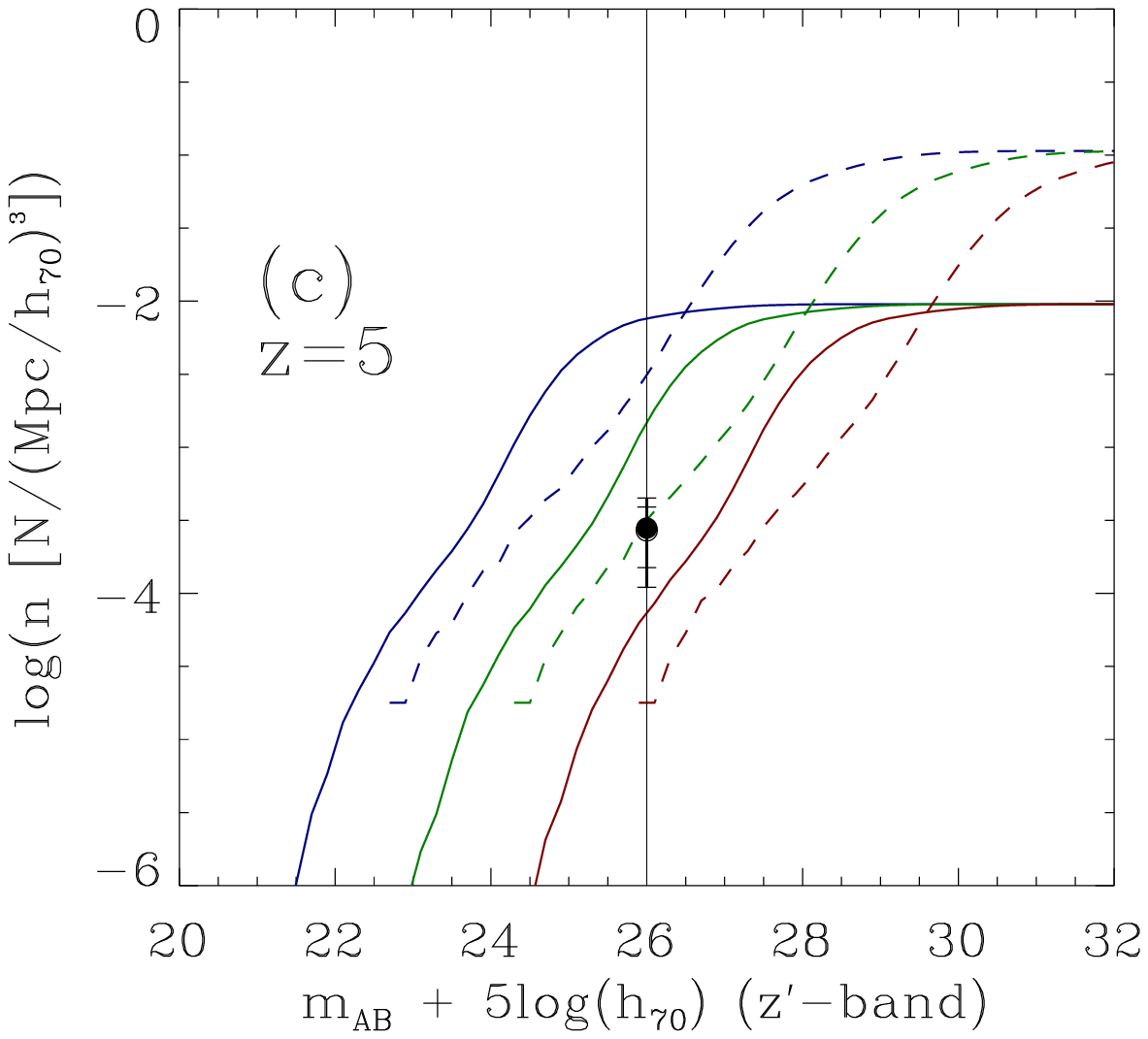}
\includegraphics[width=80mm]{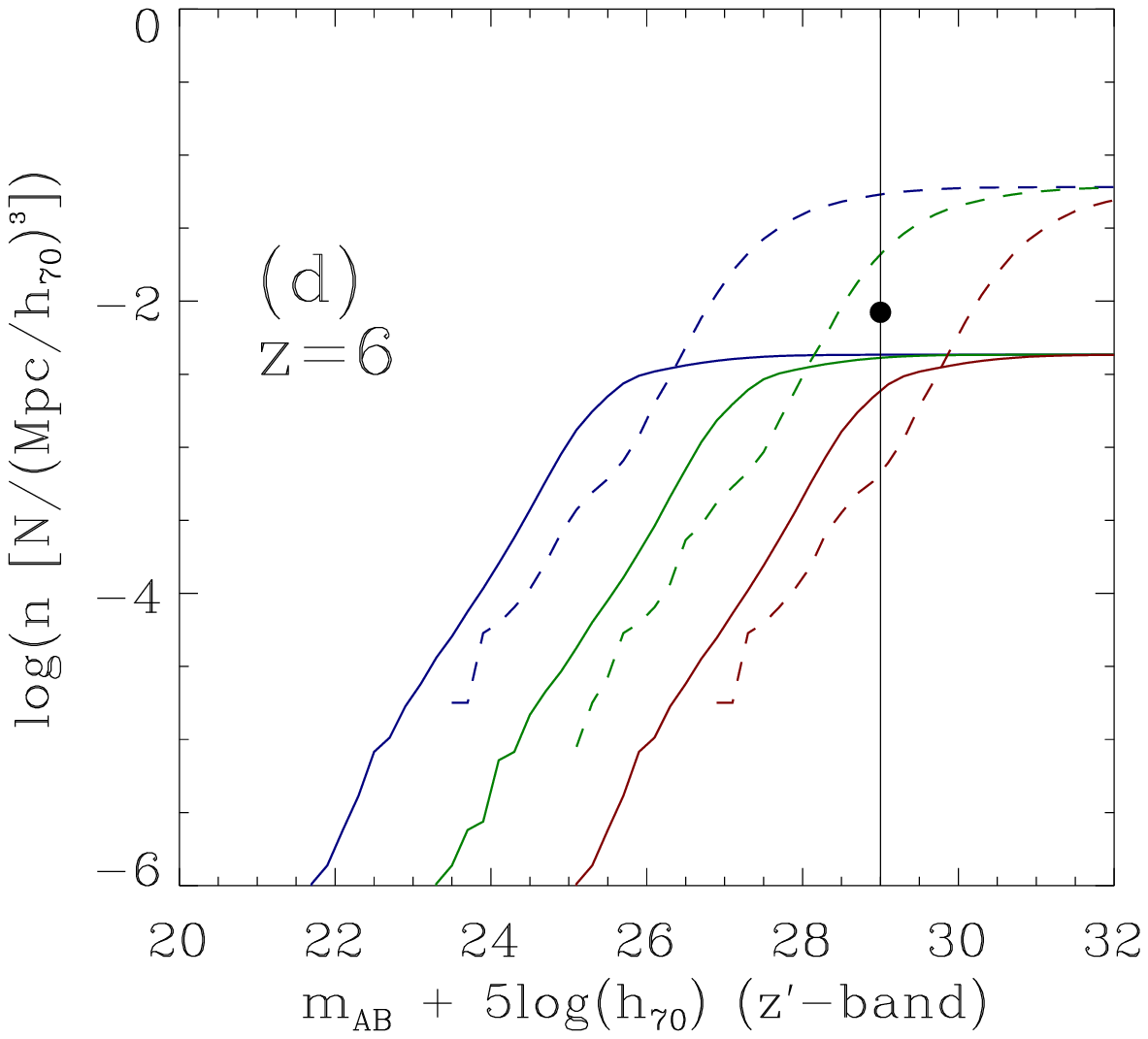}
\end{center}
\caption{Total number density of galaxies below a threshold magnitude 
  in the `Large' box size (G6 run, solid curves) and `Medium' box size (D5
  run, dashed curves) simulations. Blue, green, and red curves represent the
  extinction values $E(B-V)=0.0$, 0.15, and 0.3, respectively.  The vertical
  solid black lines roughly represent the magnitude limits of the
  \citet{Steidel03} sample for $z=3$, the SDF sample for $z=4$ and 5, and the
  HST GOODS for $z=6$. }
\label{density}
\end{figure*}

Fig.~\ref{density} shows the cumulative number density (integral of
the luminosity function for galaxies brighter than a certain magnitude
limit) of galaxies at various redshifts and for different extinction
values for the `Large' (solid curves) and `Medium' (dashed curves) box
sizes. Every simulation of limited size will underpredict the
continuum value by a certain amount, depending on the bright-end
cutoff imposed by the finite volume, so it is not surprising that the
`Medium' simulation gives systematically lower densities than the
`Large' simulation. For comparison, we show values determined by
\cite{Ade03} at $z=3$, \cite{Ouchi04a} at $z=4$ and 5, and
\cite{Bouwens04} at $z=6$. The best match with the observations is
reached for $E(B-V)=0.15$ at $z=3$, but the required extinction
appears to slightly increase towards $E(B-V)=0.3$ at higher redshifts.

At $z=3$ and for $E(B-V)=0.15$, we find a value for the number density
of $n(R<25.5) \sim 1 \times 10^{-3} (\himpc)^{-3}$. This magnitude was
determined by the limiting magnitude in the survey of \cite{Ade03};
for values at different magnitudes, refer to Fig.~\ref{density}.
Similarly, at $z=4$ and 5, we determine values of $n(i'<26.5) \sim 6
\times 10^{-3} (\himpc)^{-3}$ and $n(z'<26.0) \sim 1.5 \times 10^{-3}
(\himpc)^{-3}$. The variation in these results primarily reflects the different limiting
magnitudes of \cite{Ouchi04a} instead of an internal evolution of the
LF over this redshift range.  Finally, at $z=6$, we measure a value of
$n(z'<29.0) \sim 2.1 \times 10^{-2} (\himpc)^{-3}$, where the limiting
magnitude was chosen as in \citet{Bouwens04}.


\section{Luminosity Functions}
\label{section:lf}

The most widely used analytic parameterisation of the galaxy luminosity
function is the Schechter Function \citep{Schechter}, the logarithm of which
is given by
\begin{equation}
\log(\Phi(M)) = \log(0.4 \ln(10)\Phi^*)+\mu(\alpha+1)-10^\mu/\ln(10),
\label{eq:schechter}
\end{equation}
where $\mu=-0.4(M-M^*)$, and $\Phi^*$, $M^*$, and $\alpha$ are the
normalisation, characteristic magnitude, and faint-end slope, respectively.
We note that throughout this section, we plot luminosity functions (LFs) in
terms of magnitude rather than luminosity. Brighter objects
will thus appear farther left on the abscissa than fainter objects. 

The LF measured from our simulations suffers both from boxsize and
resolution effects, so we expect it to be physically meaningful only
for a certain limited range of luminosities. At the faint end, objects
are made up by a relatively small number of particles, and may not be
well-resolved, and even smaller objects will be lost entirely.  In our
simulations, the luminosity functions generally have a peak at around
100 stellar particles. The turn-over on the dim side of this peak 
owes to the
mass resolution.  In order to avoid being strongly affected by this
limitation, we usually discard results based on galaxies with fewer
than 200 stellar particles.

At the bright end, objects become increasingly rare. When there is
only of order one object per bin in the entire box, the statistical
error of the LF dominates and we cannot reliably estimate the
abundance. In order to improve the sampling of these objects, a larger
simulation volume needs to be chosen, which is however in conflict
(for a given particle number) with the usual desire to obtain a good
mass resolution.

\begin{figure}
\begin{center}
\includegraphics[width=84mm]{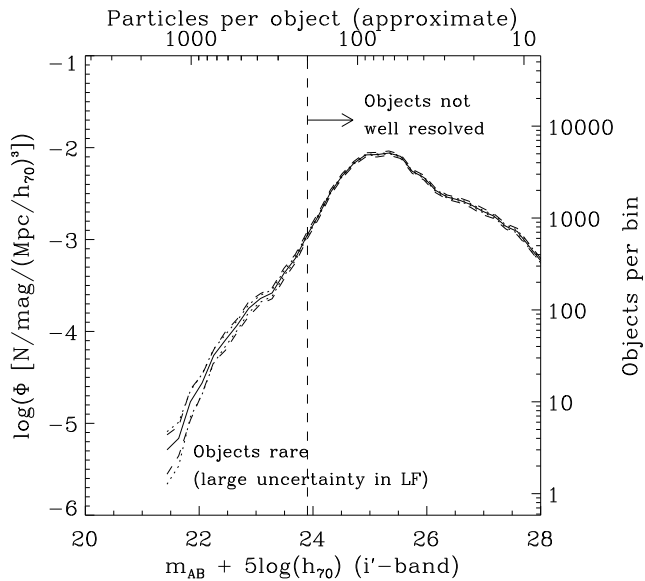}
\end{center}
\caption{Typical luminosity function, depicting our approximate
criterion for physical relevance of the data, and uncertainty. The top
and right axes show simulation-based measurements of objects: the
approximate number of stellar particles which make up an object in the
given magnitude bin, and the number of objects in the entire box in a
given magnitude bin (of size 0.2 mags).  The bottom and left axes are
the corresponding physical measurements: the magnitude as derived from
the spectrum of the objects, and the number density as a luminosity
function. A fiducial cutoff of 200 stellar particles is chosen as the
lower limit for accurately resolved objects, although this value is
somewhat arbitrary. The dotted line indicates Poisson $\sqrt{N}$
uncertainties, and the dashed line indicates the standard deviation of
the mean between the luminosity function derived for eight disjoint
sub-regions of the entire box. Since the latter uncertainty is greater
throughout most of the range, only it is shown on subsequent
plots. Both forms of uncertainty are dependent on the bin size, 0.2
mags.}
\label{LF-example}
\end{figure}

Fig.~\ref{LF-example} illustrates these numerical limitations and
defines a region where the LF results can be trusted.  The cut-offs we
indicated here are however approximate.  A precise determination of
the interval over which the measured LF is physically significant
requires running many simulations with successively higher resolution,
and looking for an interval of convergence. Such a programme was
carried out by \citet{SH03b} in their study of the cosmic star
formation rate.  We here analyse three different box sizes
taken from their set of simulations, yielding a good coverage of
magnitude space. Where appropriate, we also combine these results to
obtain a measurement covering a larger dynamic range.  In subsequent
figures, the approximate interval of physical significance for the LF
measurements will be shown with a thicker line than the rest of the
curve.  Strictly speaking, it is only this interval that can be
compared reliably with observations.

\begin{figure}
\begin{center}
\includegraphics[width=84mm]{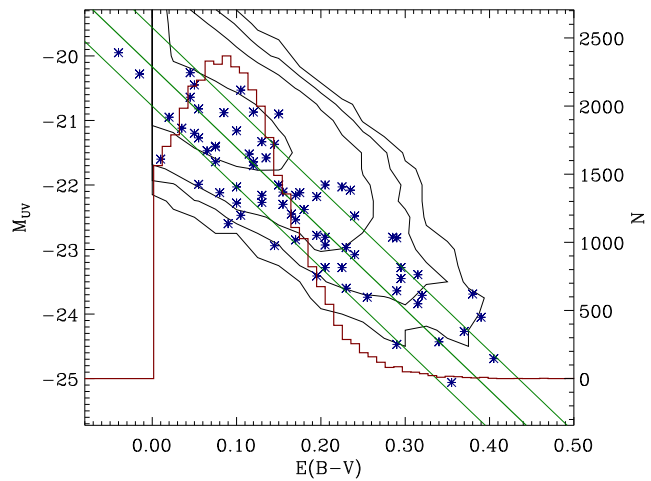}
\end{center}
\caption{Correlation between UV magnitude and extinction value
$E(B-V)$. Blue points, plotted with respect to the left axis, show the
original data from \citet{Shapley}. Green lines show the best linear
fit to the data, and the lines denote 1-$\sigma$ scatter. Black
contours outline the extinction values we assigned to the simulated
galaxies using the procedure described in Section~\ref{section:lf}
(for the `Large' boxsize G6-run at $z=4$) in order to match up with
this correlation. The red histogram, plotted with respect to the right
axis, shows the distribution of extinction values of the black
contours. However, for the reasons described in the text, we do not
use this method of assigning variable $E(B-V)$ hereafter.}
\label{E-Shapley}
\end{figure}

Also shown in Fig.~\ref{LF-example} is the estimated uncertainty in
the LF owing to sample size, which we estimated in two ways. First, we
calculated $\sqrt{N}$ Poisson statistical errors, simply by taking the
square root of the number of objects in each bin. Second, we divided
the box into eight octants, and computed the standard deviation of the
mean between the LFs computed with each of the eight octants. In all
cases, the second method produced larger uncertainty over the interval
of physical significance, indicating that the `cosmic variance' error
owing to our limited boxsize exceeds a simple Poisson estimate. We
therefore use error estimates obtained with the octant method in our
subsequent figures on the LF results and ignore the Poisson errors.

Typically, we assumed three values for the extinction, $E(B-V)=0.0$,
0.15, and 0.30, and produced LF-estimates for them separately. In this
procedure, we hence always assigned a single extinction value to every
galaxy for which we computed magnitudes.  However, in the real
universe we instead expect a distribution of extinction values
\citep[e.g.,][]{Shapley, Ouchi04a}, which could be quite broad. This
prompted us to explore possible effects owing to a `variable
extinction'.  To this end, we first introduced random scatter into
the values for extinction: Instead of appling the same extinction to
all galaxies, each galaxy was assigned an individual value of $E(B-V)$
determined by a Gaussian random variable with a mean of 0.15 and a
standard deviation of 0.10.  Moreover, a cutoff was imposed so that no
galaxy had a negative extinction value. We found that such variable
scatter tends to smooth out the luminosity function somewhat, as
expected, but it does not produce results readily distinguishable from
a uniform extinction value.
\begin{figure*}
\includegraphics[width=80mm]{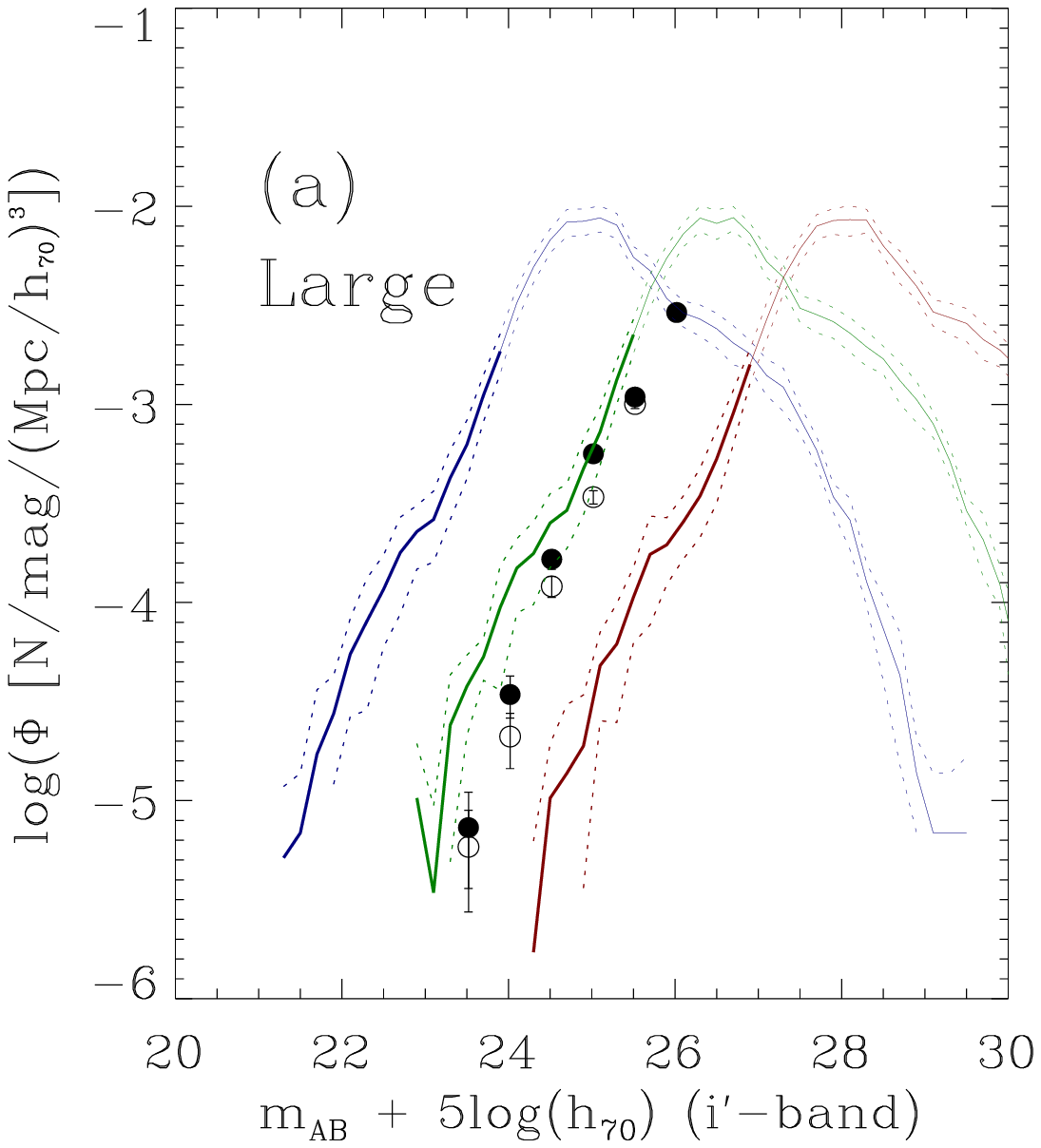}
\includegraphics[width=80mm]{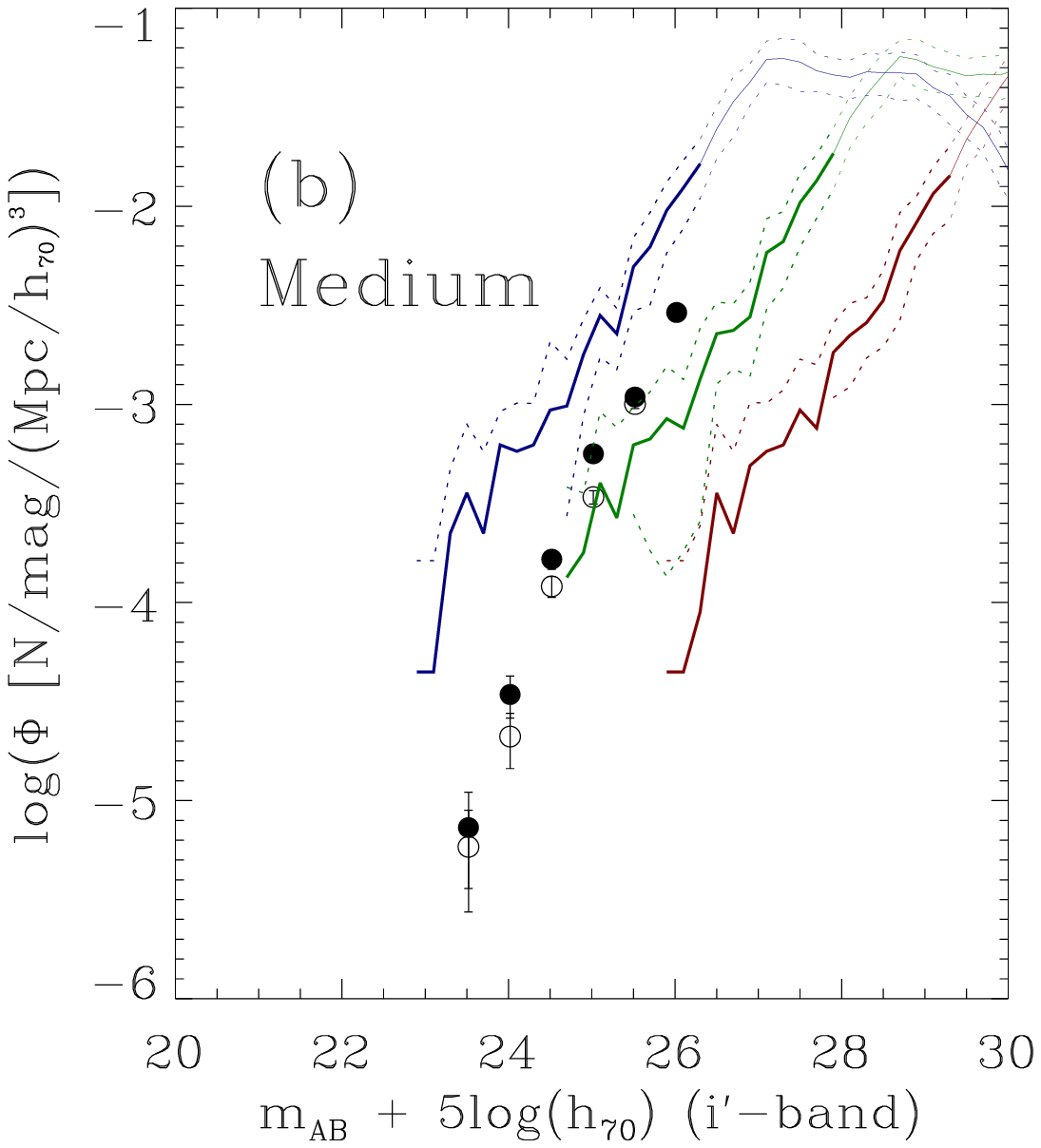}
\caption{$i'$-band luminosity function for both `Large' 
  (G6, panel a) and `Medium' boxsize (D5, panel b) simulation runs at $z=4$,
  and the observational data for $BRi$--LBGs \citep[][Fig.  16]{Ouchi04a}.
  Simulation LFs are plotted as blue, green, and red curves representing
  extinction values of $E(B-V)=0.0$, 0.15, and 0.30, respectively. The Subaru
  survey data are shown as black crosses and boxes for the two different
  survey fields \citep[][Fig. 16]{Ouchi04a}.  }
\label{LF-z4-data}
\end{figure*}

\begin{figure}
\begin{center}
\includegraphics[width=84mm]{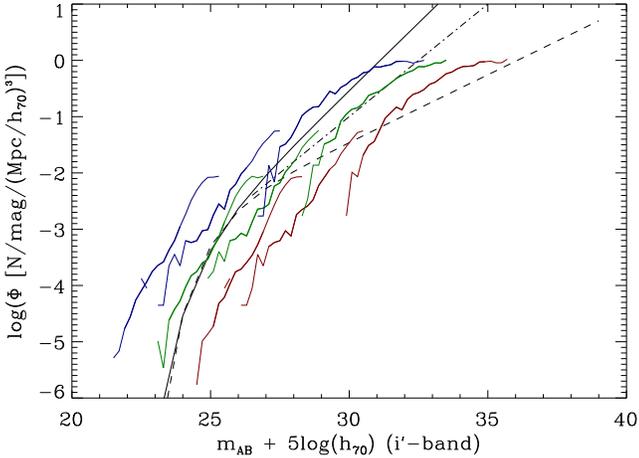}
\end{center}
\caption{Luminosity function for all three boxsizes 
  at $z=4$. Extinction colour-coding is the same as in the previous plots.
  Solid and dashed lines are best Schechter fits to Subaru $BRi$--LBG data,
  assuming a value for the faint-end slope of $\alpha= -2.2$ and -1.6, 
  respectively \citep[][Fig. 16]{Ouchi04a}. The dashed-dotted line has 
  a slope of -2.0, but is not a fit to the Subaru data.}
\label{LF-GDQz4-line}
\end{figure}

There is evidence for a correlation between UV magnitude and
extinction.  In particular, \citet{Shapley} found that, over the
magnitude range studied, dust obscures almost exactly enough flux to
give all galaxies a similar apparent magnitude. The data from
\cite{Shapley}, along with the extinction values we adopted to emulate
it, appear in Fig.~\ref{E-Shapley}. When these extinction values are
used, brighter galaxies have larger values of $E(B-V)$.  The effect of
this distribution of extinction values is such that the brightest
galaxies become dim enough to agree with the observed magnitudes,
while the fainter ones suffer little extinction so that they agree
with the narrow range of observed magnitude.  Taken together, this
effect produces a simulated LF that matches the empirical one quite
well within the observed range of magnitudes, requiring only moderate
extinction at the faint end. However, invoking variable extinction in
this manner is of course bound to succeed at some level, because we
here `hide' the difference between simulated and observed LFs in the
variable extinction law. While such a law in principle may exist, we
prefer here to systematically study the effect of extinction by
assuming different values of $E(B-V)$ uniformly for the entire sample.

\begin{figure*}
\includegraphics[width=80mm]{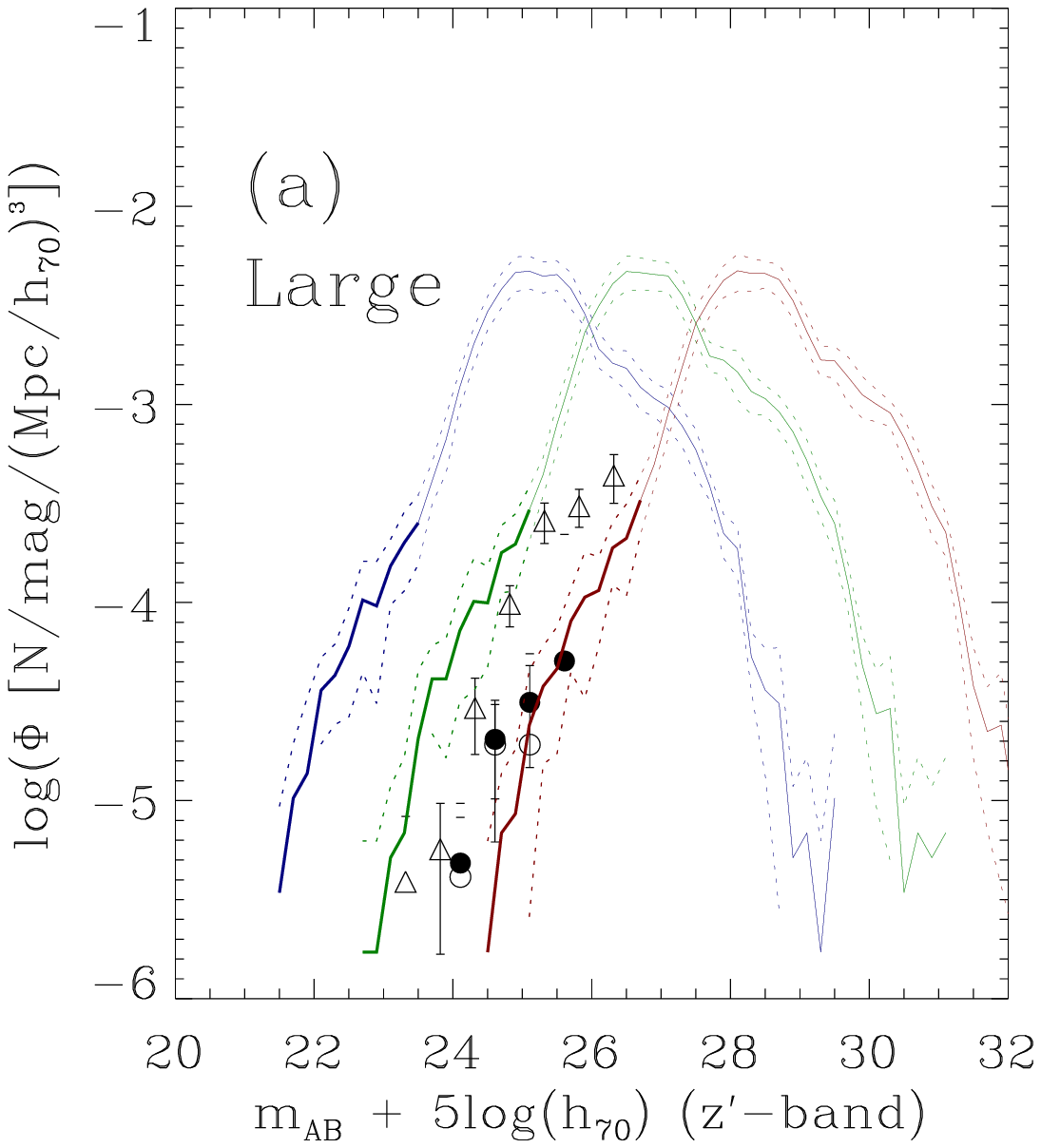}
\includegraphics[width=80mm]{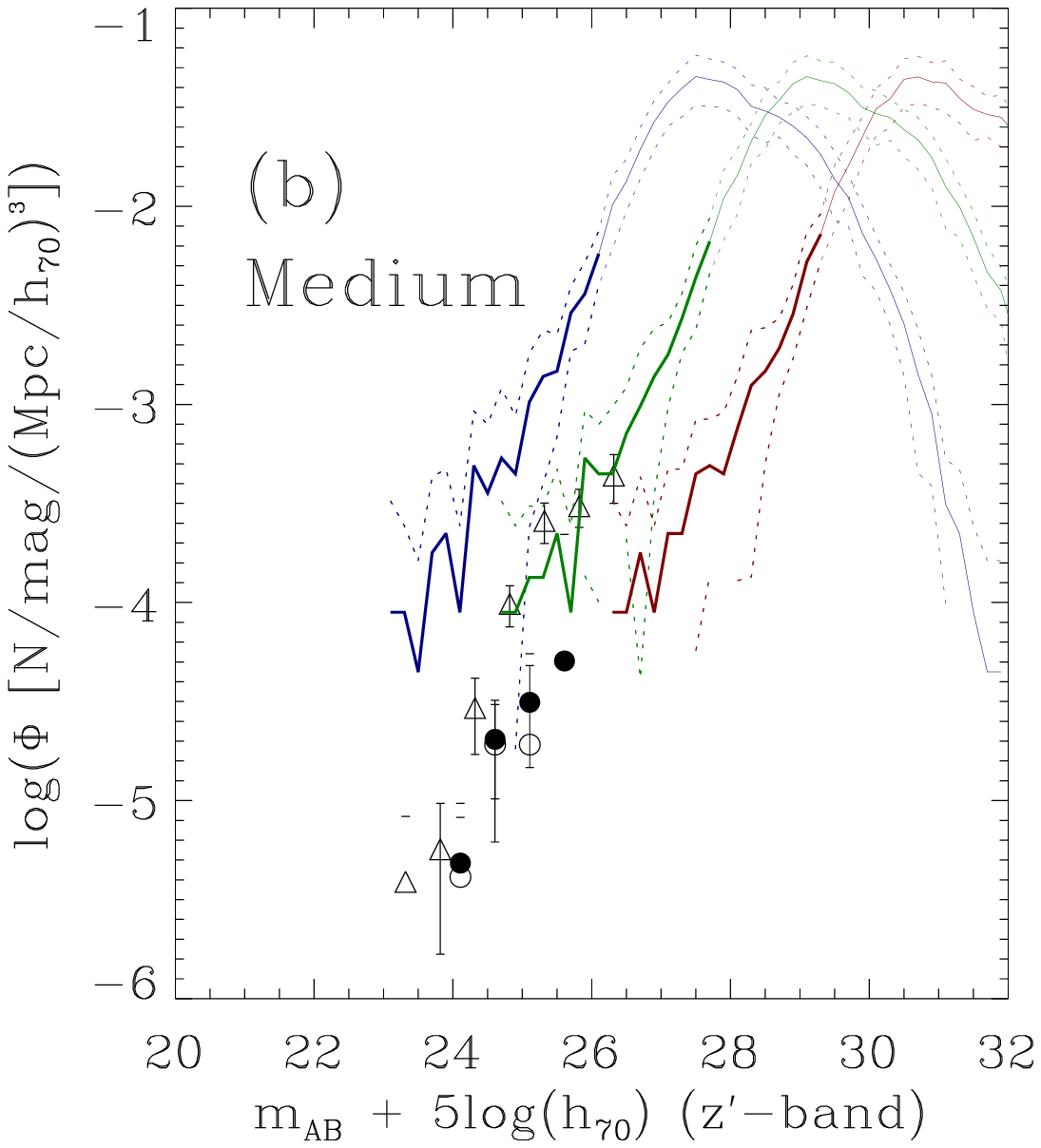}
\caption{$z'$-band luminosity function for both `Large' 
  (G6, panel a) and `Medium' boxsize (D5, panel b) simulations at $z=5$.
  Survey results for similar redshifts are shown with symbols. Simulation data
  are plotted as blue, green, and red curves representing extinction values of
  $E(B-V)=0.0$, 0.15, and 0.30, respectively. Ouchi et al.'s  $z'$-band survey
  data are shown as black crosses and boxes for the two different survey
  fields  \citep[][ Fig. 16]{Ouchi04a}.  The updated $I$-band survey data of
  \citet{Iwata04} are shown as triangles.  }
\label{LF-z5-data}
\end{figure*}

\begin{figure}
\begin{center}
\includegraphics[width=84mm]{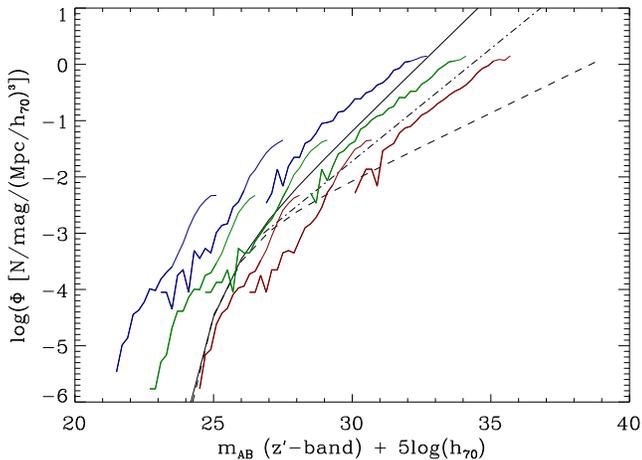}
\end{center}
\caption{$z'$-band luminosity function for all three 
  boxsizes at $z=5$. Extinction colour-coding is the same as in the previous
  plots. Solid and dashed lines are best fit Schechter functions to Subaru
  $Riz$--LBG data, assuming a value for the faint-end slope of $\alpha= -2.2$
  and $-1.6$, respectively \citep[][ Fig. 16]{Ouchi04a}. The dashed-dotted
  line has a slope of $-2.0$, but is not a fit to the data.}
\label{LF-GDQz5-line}
\end{figure}

\begin{figure*}
\includegraphics[width=80mm]{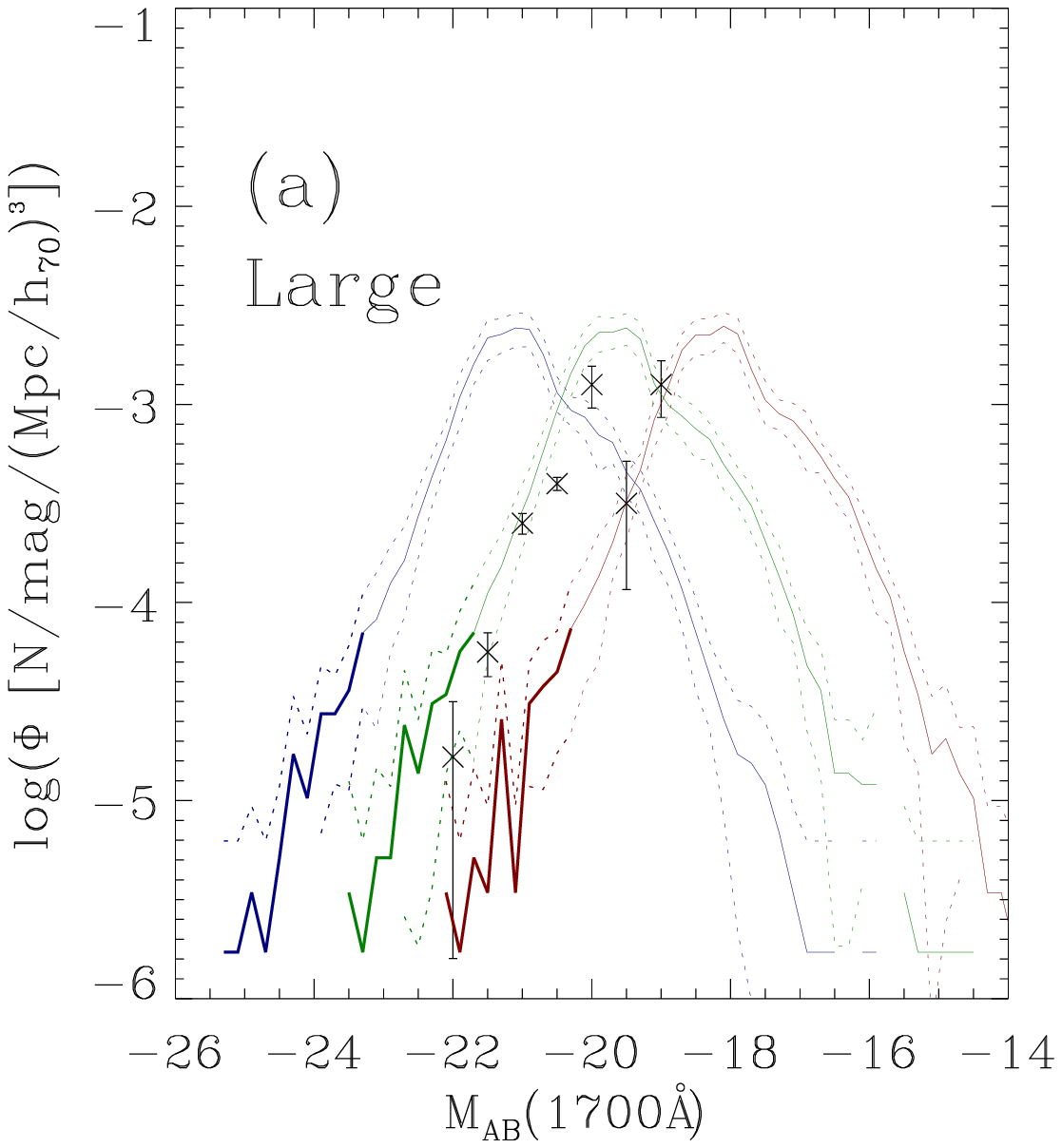}
\includegraphics[width=80mm]{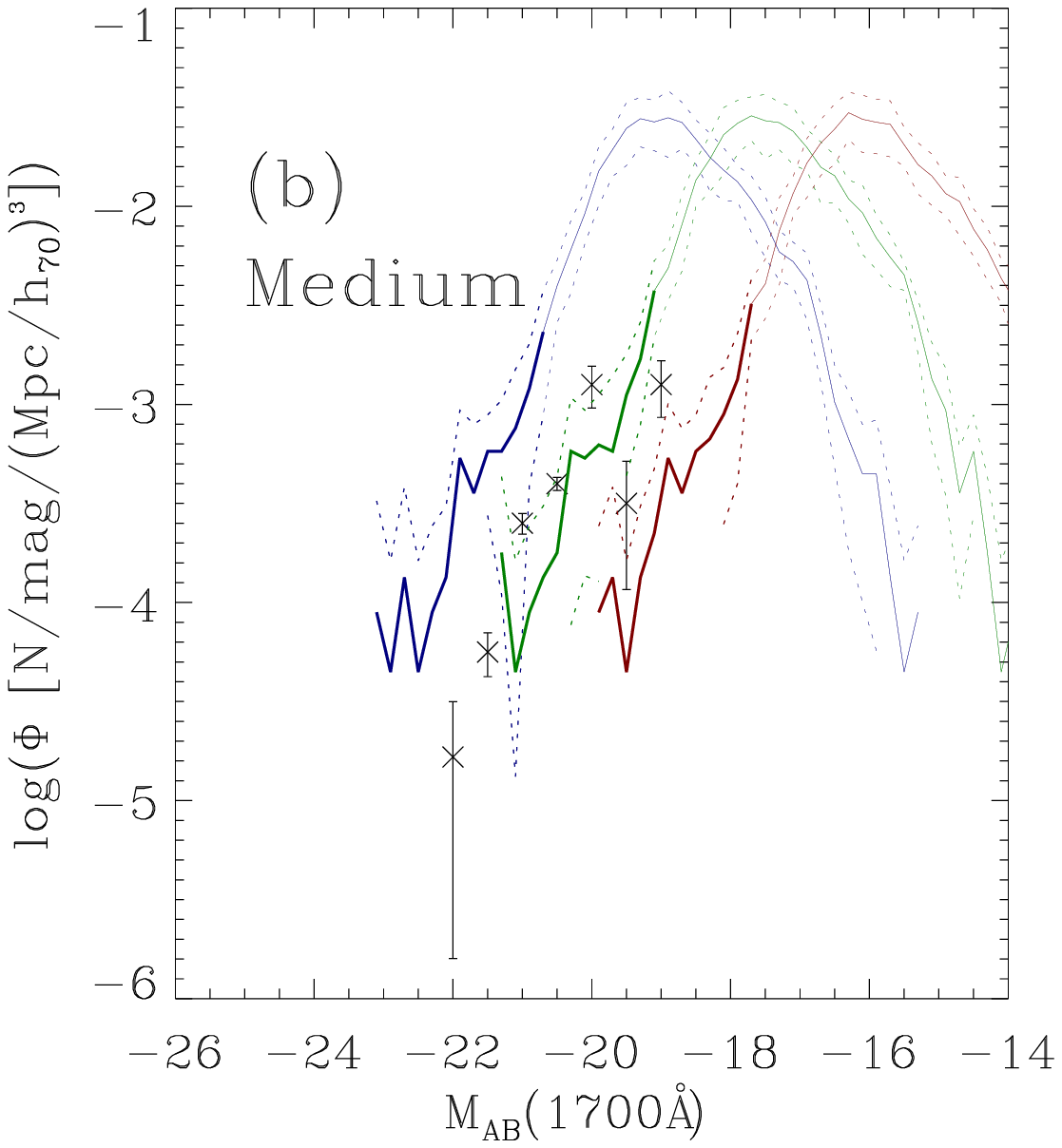}
\caption{UV luminosity function at $z=6$ for 
  both `Large' and `Medium' boxsize simulations and the observational data
  from \citet[][ black crosses]{Bouwens04}.  Simulation data are plotted as
  blue, green, and red curves, representing extinction values $E(B-V)=0.0$,
  0.15, and 0.30, respectively.  }
\label{LF-z6-data}
\end{figure*}

In Figs.~\ref{LF-z4-data} through \ref{LF-z6-data}, we show the LFs
derived from the simulated observations as coloured curves with data
points from observational surveys overplotted.  These are the main
results of this paper.  We first compare the $i'$-band LFs directly to
the observations. For $z=4$, this is shown in
Fig.~\ref{LF-z4-data} for the `Large' (G6) and `Medium' boxsize (D5)
runs, along with data from \cite{Ouchi04a}. The observational data
points do not extend faint enough to offer any overlap with the
`Small' (Q5 \& Q6) run's LF, therefore we do not show the results from
the `Small' simulation in this figure. As seen in the figures, the
data points are roughly consistent with the green curves,
corresponding to an extinction value of $E(B-V)\sim 0.15$. The result
of our `Large' run covers the entire range of the observed magnitude
ranges, but the `Medium' run overlaps only with the fainter side of
the Subaru data because of the moderate simulation volume.

We then compared the simulated LFs for all three boxsizes (`Large',
`Medium', and `Small') to the best-fitting Schechter function as
determined from the Subaru survey. For $z=4$, this is shown in
Fig.~\ref{LF-GDQz4-line}. The two Schechter functions plotted in
this figure correspond to the best fit to the Subaru data obtained by
\citet{Ouchi04a} with a fixed faint-end slope of either $\alpha =
-1.6$ (dashed line) or $-2.2$ (solid line). The simulated LF with
$E(B-V)=0.15$ is consistent with the faint-end slope between these two
values, and the value of $\alpha=-2.0$ (dash-dotted line) appears 
reasonable.

Similar results emerged for $z=5$. In Fig.~\ref{LF-z5-data}, we show
the `Large' and `Medium' boxsize LFs and compare them with data from
\cite{Ouchi04a} and \cite{Iwata04}. At $z=5$, the survey data appear
to be more consistent with the simulations using $E(B-V)=0.3$,
particularly at the bright end, suggesting a slightly larger
extinction at $z=5$ than at $z=4$.  And again, when all three boxsizes
are plotted along with the best-fitting Schechter function in
Fig.~\ref{LF-GDQz5-line}, the most consistent value for the faint
end slope $\alpha$ appears to lie between $-1.6$ and $-2.2$.  Note
that there are small discrepancies between the LF-estimates by
\citet{Ouchi04a} and \citet{Iwata04}.  This may owe to slight
differences in the colour-selection criteria used in the two surveys.

In Fig.~\ref{LF-z6-data}, we compare the UV-magnitude LFs at $z=6$ in
the `Large' and `Medium' boxsize simulations to data from
\cite{Bouwens04} .  Again, a value of $E(B-V)$ between 0.15 and 0.30
leads to the best match with the data, and an extinction with
$E(B-V)=0.15$ is favoured based on this comparison.

\begin{figure}
\begin{center}
\includegraphics[width=84mm]{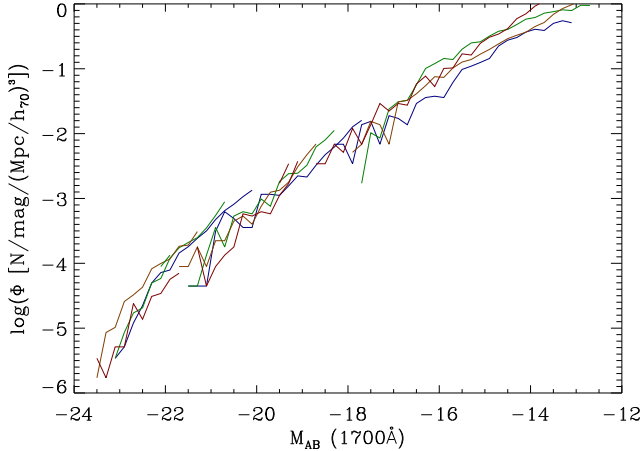}
\end{center}
\caption{Rest frame UV luminosity function for all 
  three boxsizes for redshifts $z=6$ (red), 5 (orange), 4 (green), and 3
  (blue). For clarity, results for each LF are only plotted over the reliable
  range, and error bars are omitted. The extinction value for all curves is
  here $E(B-V)=0.15$.}
\label{LF-GDQz6543}
\end{figure}

Finally, Fig.~\ref{LF-GDQz6543} examines the evolution of the LF
over the redshift range studied.  To this end, we plot all redshifts
for all boxsizes using the single extinction value $E(B-V)=0.15$.
Overall, the LF of our simulations shows little if any evolution over
the redshift range in question.  The absence of strong evolution is
probably related to the fact that the evolution of the cosmic star
formation rate (SFR) density is quite mild from $z=3$ to 6 in our
simulations, as discussed by \citet{SH03b}
and Hernquist \& Springel (2003).  In both SPH and total
variation diminishing (TVD) simulations \citep{Nachos1}, the cosmic
SFR continues to rise gradually from $z=3$ to 5, and peaks at $z=5-6$.
\citet{Nachos2} have shown that the evolution of the LF from $z=3$ to
$z=2$ is about 0.5 mag, so it is perhaps not too surprising that the
evolution at higher redshift is of comparably small size. Note that in
terms of proper time, the redshift interval from $z=6$ to 3 is only
about as long as the interval from $z=3$ to 2.


\section{Discussion \& Conclusions}
\label{section:conclusion}

Using state-of-the-art cosmological SPH simulations, we have derived
the colours and luminosity functions of simulated high-redshift
galaxies and compared them with observations. In particular, we have
employed a series of simulations with different boxsizes and
resolution to identify the effects of numerical limitations. We find
that the colours of galaxies at $z=4-6$ agree with the observed ones
on the colour-colour planes used in observational studies, and 
our results confirm the generic conclusion from earlier numerical studies
\citep{Nag02, Wei02, NSHM, Nachos2} that UV bright LBGs at $z\ge 3$ are 
the most massive galaxies with $E(B-V)\sim 0.15$ at each epoch. 

The simulated LFs are in good agreement with the data provided an
extinction of $E(B-V)=0.15 - 0.30$ is assumed. The faint-end slope of
our results is consistent with a value between $\alpha=-1.6$ and
$-2.2$, as found with the Subaru data \citep{Ouchi04a}. The simulated
LFs best match a very steep faint-end slope of $\alpha \sim -2.0$. The
recent analysis of {\it Hubble Space Telescope} Ultra Deep Field (UDF)
data by \citet{Yan04} also suggests a quite steep faint-end slope of
$\alpha=-1.8$ to $-1.9$, in good agreement with the simulations.

The steep faint-end slopes at $z\ga 6$ found here are interesting
because they have significant implications for the history of the
reionisation of the Universe.  As discussed by, e.g.,
\citet{SH03b}, 
\citet{Yan04}, and
\citet{Nachos4}, 
it may be possible to reionise the
Universe at $z=6$ with ionising photons from Population II stars in
normal galaxies alone if the faint-end slope of the galaxy luminosity
function is sufficiently steep ($\alpha \la -1.6$).  Note that
\citet{NSHM} also found a similarly steep faint-end slope at $z=3$ in
the same SPH simulations.

The suggestion that the faint-end slope may be much steeper at high
redshifts than in the Local Universe is intriguing, and the agreement
between the recent observations and our simulations lends encouraging
support for this proposal.  However, this immediately invokes the
question of what mechanisms changed the LF over time and gave it the
shallow slope ($\alpha\sim -1.2$) observed locally in the 2dF
\citep{Cole01} and SDSS surveys \citep{Blanton01}.  A similarly strong
flattening does not occur in our simulations \citep{NSHM}, but this
could owe to an incomplete modeling of the physics of feedback
processes from supernova and quasars.  We note that the SPH
simulations employed in this paper already included a galactic wind
model \citep[see][for details]{SH03a} which drives some gas out of
low-mass halos, but the effect is not sufficiently sensitive to galaxy
size to produce a significantly flattened faint-end slope. The
shallowness of the faint-end slope of the local LF hence remains a
challenge for cosmological simulations and may point to the need for
other physical processes, such as black hole growth
\citep[e.g.,][]{Springel2005b, DiMatteo2005, Robertson2005}.

In light of this, it is therefore particularly encouraging that the
simulations are more successful at high redshift. The steep faint-end
slope found by observations in this regime is an important constraint
on the nature of the feedback processes themselves. Presently, the
observational results still bear substantial uncertainties, however,
stemming mostly from their limited survey volume which makes them
prone to cosmic variance errors, and less from the faintness levels
reached, which already probe down to $m_{AB}=29$ magnitude in the case
of the UDF survey.  The next generation of wide {\em and} deep surveys
will shed more light on this question in the near future.


\section*{Acknowledgments}

We thank Ikuru Iwata for useful discussions and providing us with the
unpublished luminosity function data. We are also grateful to Masami
Ouchi for providing us with the LF data and the Subaru filter
functions, and to Alice Shapley for providing us with the extinction
data. This work was supported in part by NSF grants ACI 96-19019, AST
98-02568, AST 99-00877, and AST 00-71019.  The simulations were
performed at the Center for Parallel Astrophysical Computing at the
Harvard-Smithsonian Center for Astrophysics.


\end{document}